\documentclass[reprint,prd,twocolumn,superscriptaddress,showpacs,nofootinbib,floatfix,
preprintnumbers]{revtex4-1}
\usepackage{float}
\usepackage{color}
\usepackage{hyperref}
\usepackage{graphicx, bm}
\def\lsim{\mathrel{\rlap{\lower4pt\hbox{\hskip1pt$\sim$}}
    \raise1pt\hbox{$<$}}}  
\def\gsim{\mathrel{\rlap{\lower4pt\hbox{\hskip1pt$\sim$}}
    \raise1pt\hbox{$>$}}}                

\usepackage{amssymb}
\usepackage{amsmath}

\begin{document}

\title {\bf Non-perturbative over-production of axion-like-particles (ALPs) via derivative interaction} 

\author{Anupam Mazumdar} 
\affiliation{Consortium for Fundamental Physics, Physics Department, 
Lancaster University, Lancaster LA1 4YB, United
  Kingdom}
\affiliation{Institute for Particle Physics Phenomenology,
Durham University, Durham DH1 3LE, UK} 
\author{Saleh Qutub} 
\affiliation{Consortium for Fundamental Physics, Physics Department, 
Lancaster University, Lancaster LA1 4YB, United
  Kingdom}
\affiliation{Department of Astronomy, King Abdulaziz University, Jeddah 21589, Saudi Arabia}
\vspace{1cm}


\begin{abstract}
Axion like particles (ALPs) are quite generic in many scenarios for physics beyond the Standard Model, they are pseudoscalar Nambu-Goldstone bosons, and 
appear once any global $U(1)$ symmetry is broken spontaneously. The ALPs can gain mass from various non-perturbative quantum effects, such as anomalies or instantons. ALPs can couple to the matter sector incluidng a scalar condensate such as inflaton or moduli field via derivative interactions, which are suppressed by the axion {\it decay constant}, $f_\chi$ . Although weakly interacting, the ALPs can be produced abundantly from the coherent oscillations of a homogeneous condensate. In this paper we will study such a scenario where the ALPs can be produced abundantly, and in some cases can even overclose the Universe via odd and even dimensional operators, as long as $f_\chi/\Phi_{\rm I} \ll 1$, where $\Phi_{\rm I}$ denotes the initial amplitude of the coherent oscillations of the scalar condensate, $\phi$. 
We will briefly mention how such dangerous overproduction would affect dark matter and dark radiation abundances in the Universe.
\end{abstract}

\maketitle


\section{Introduction}

 Inflation~\cite{inflation} is extremely  successful  paradigm for creating seed perturbations for the large scale structure in the Universe~\cite{Planck:2015xua}. One of the post-inflationary challenges is to recreate the Universe we live in - filled with the Standard Model degrees of freedom and the dark matter particles~\cite{Albrecht:1982mp}.   The energy extraction from inflaton energy density to matter degrees of freedom is non-trivial, and it all depends on the couplings, i.e how inflaton couples to other matter degrees of freedom, for a review, see~\cite{review1,review2}.

Similar to the inflaton, there could be many light moduli fields present in the early Universe, whose potentials are absolutely flat. 
The moduli fields are natural consequence of any low energy string theory, which appear in $4$ dimensions with almost flat potential~\cite{Nilles}. In any supersymmetric 
theory there would be flat directions, which are built upon $F$ and $D$ flatness condition, see~\cite{MSSM}.
 Typically, the moduli could be made heavy by stringy non-perturbative effects~\cite{Giddings:2001yu}, but still there could be one light moduli with a mass around the scale of low supersymmetry (SUSY) breaking. If the low scale SUSY breaking is at ${\cal O}(\rm TeV)$, then the moduli would typically take a mass of that order. Light moduli could be displaced during inflation by virtue of quantum fluctuations of order the Hubble expansion rate during inflation, see~\cite{Starobinsky:1986fx,Linde}. In which case, the moduli lighter than the Hubble expansion rate during inflation could be settled with a large vacuum expectation value (VEV), i.e. of the order of 
 $M_P\sim 2.4\times 10^{18}$~GeV. The moduli behaves like a condensate~\cite{MSSM}, and begins its coherent oscillations when the Hubble expansion rate of the Universe drops to the mass of the moduli~\footnote{In principle one should check the flatness condition of the moduli or flat direction potential by taking various SUSY breaking inputs in a time dependent background~\cite{Allahverdi:2008bt}.}.
 
It has been recognised, sometime ago, that the energy transfer from the inflaton/moduli field to its decay products need not be perturbative. The perturbative decay is typically slow, and the time scale is determined by the total decay width of the inflaton/moduli. On the other hand, there is an interesting possibility of draining the energy density of inflaton/moduli via non-perturbative effects in a time dependent background - known as preheating~\cite{Preheating}, and \cite{Kofman:1994rk,Kofman:1997yn}.
 
 The particle production during preheating is indeed non-adiabatic, the effective time dependent frequency does not evolve adiabatically. The 
 violation of adiabaticity condition is the key for particle creation~\cite{Kofman:1997yn}. 
 Another important non-perturbative production mechanism, dubbed as tachyonic preheating, takes place when the time dependent frequency becomes imaginary which happens for instance when symmetry is broken spontaneously~\cite{Felder:2000hj,Felder:2001kt}, or when the trilinear coupling comes to dominate the interaction with the inflaton/moduli~\cite{Dufaux:2006ee}.
 The process  of particle creation, soon becomes non-linear due to the mode-mode interactions, which makes the problem soon intractable. The epoch of preheating, or explosive particle production stops  when the back reaction is taken into account~\cite{Kofman:1997yn,Micha:2002ey,Felder:2000hr}. Nevertheless, during the first few oscillations, as long as the energy density in the decayed quanta is still below the energy density stored in the inflaton/moduli sector, we can trust the analytical treatment, see for instance~\cite{Cormier:2001iw}.

During preheating one can excite scalars~\cite{Preheating,Kofman:1994rk,Kofman:1997yn}, fermions~\cite{Greene:1998nh}, higher spins~\cite{Maroto:1999ch} and gauge fields~\cite{Gauge,Mazumdar:2008up,Adshead:2015pva}. Preheating can also generate isocurvature perturbations~\cite{metric}, and non-Gaussian signatures~\cite{Enqvist:2005qu,Enqvist:2004ey}. One can also excite fairly long-lived dark matter particles~\cite{Kofman:1997yn,dark}. In fact, the large abundance of dark matter particle created at the time of preheating is the real threat for cosmology. 

In this paper we are keen to understand the cosmological consequences of inflaton or moduli interaction via {\it derivative} couplings.
For simplicity we will not distinguish between inflaton or moduli field, we will call the field to be a generic {\it scalar condensate} without loss of generality, 
which {\it oscillates} coherently when the scalar condensate mass becomes equal to that of the Hubble expansion rate. 

One such example of derivative interaction is demonstrated by 
{\it axion-like} particles (ALPs) defined here as $\chi$, for a review, see~\cite{Ringwald:2014vqa}.  The ALP appears due to the spontaneous breaking of 
a global $U(1)$ symmetry.  At low energies  below the symmetry breaking scale, a Nambu-Goldstone boson arises as massless excitations of the angular part of the 
complex scalar field, which is an absolute gauge singlet. The ALPs can couple to the scalar field via dimensional $5$ and higher dimensional operators. The 
 operators are suppressed by the axion decay constant, $f_\chi \leq M_P$, in $4$ dimensions. Typically $f_\chi$ could be large close to the GUT scale or
 the string scale, or even the Planck scale. We will keep $f_\chi$ as a free parameter. 

Typically, the axion such as in the case of QCD receives its mass from instanton~\cite{Peccei:1977hh,Weinberg:1977ma,Wilczek:1977pj}, in the string theory context 
the anomalous $U(1)$ gets broken non-perturbatively by the world-sheet instantons, and its mass could be large or small depending on the action of the 
instanton~\cite{Svrcek:2006yi}. 

ALPs can act as a perfect dark matter candidate, due to its weak interaction with baryonic matter field~\cite{axion-dm-0,axion-DM,axion-recent}.  On the other hand, axion produced from the decay of a scalar condensate can give rise to a non-thermal distribution of dark matter and dark radiation, depending on the mass of ALPs. Of course, one has to make sure that ALPs are produced perturbatively/non-perturbatively to match the current observations from  the Big Bang Nucleosynthesis (BBN)~\cite{BBN}, and the current Planck data on dark matter and dark radiation abundance~\cite{Planck:2015xua}~\footnote{String theory motivated models of inflation face a stern challenge from the axion over production from numerous hidden sectors, first raised in this paper~\cite{Cicoli:2010ha}. The question then arises - how to arrange various mass scales in string compactification models, such that a successful reheating without axion domination can take place~\cite{Cicoli:2010yj}.}.

The goal of this paper is to demonstrate that:\\

{\it the ALPs can be produced abundantly from the coherent oscillations of a scalar condensate for a wide range of masses, i.e. heavy or light ALPs,  
 as long as $f_\chi < \Phi_{\rm I}$, where $\Phi_{\rm I}$ denotes the initial amplitude of the coherent oscillations of a scalar condensate, i.e. inflaton/moduli.}\\

For the purpose of illustration, we will fix the initial value of the amplitude of the oscillations of the scalar condensate to be $M_P$, without loss of generality. What matters here is the ratio: $f_\chi/\Phi_{\rm I}$.  The rate at which the particle production will take place would be sensitive to this value, as we shall demonstrate in this paper. The results of this paper will have an important consequence for string theory motivated ALPs~\cite{Svrcek:2006yi,Cicoli:2010ha,Cicoli:2012aq,Higaki:2012ar}.

The paper is structured as follows - in section~\ref{sect:interactions},
we will discuss the ALP's interaction with a homogeneous scalar condensate, i.e. inflaton/moduli. In section~\ref{sec:dim5}, 
we study the ALP scalar interaction via dimensional $5$ operator, and study preheating of ALPs without and with expansion effects. We will also discuss two 
regimes: when $f_\chi\sim {\cal O}(M_P)$ and when $f_\chi < M_P$.
In section~\ref{sec:dim6},
we will discuss the ALP's interaction via dimensional $6$ operator in two regimes: when $f_\chi\sim {\cal O}(M_P)$ and when $f_\chi < M_P$. First we will 
discuss the production of ALPs, without expansion, for light, $m_\chi\leq m_\phi$, and heavy, $m_\chi> m_\phi$, ALPs. We will repeat the computation by
taking into account of the expansion effect. Finally, we will conclude in section~\ref{sec:conclusion}
 by alluding to some of the dark matter and dark radiation bounds, whose detailed scans will be provided in a separate 
publication.

\section{Interaction and equations of motion}\label{sect:interactions}

Let us consider the following Lagrangian with derivative couplings of the form~\footnote{ For the purpose of illustration, we will 
consider only dimensional 5 and 6 operators, in fact one can discuss any higher order operators, whose behaviour will be  similar to the one discussed below.}
\begin{eqnarray}
{\cal L} \ &=&
\ \frac{1}{2}\partial_\mu \phi \partial^\mu \phi - \frac{1}{2} m_\phi^2 \phi^2
+ \frac{1}{2}\partial_\mu \chi \partial^\mu \chi - \frac{1}{2} m_\chi^2 \chi^2
\nonumber \\
&& + \frac{1}{2 f_{\chi}} \phi \partial_\mu \chi \partial^\mu \chi
+ \frac{1}{4 f_{\chi}^2} \phi^{2} \partial_\mu \chi \partial^\mu \chi + \ldots \, ,
\label{lag1}
\end{eqnarray} 
where $\phi$ is a scalar condensate, which sets its coherent oscillations around the minimum of a quadratic potential governed by its mass 
$m_\phi$~\footnote{One can relax this condition and have oscillations around any kind of potential, but for many examples, such as moduli oscillations, it is a fairly
good approximation to consider coherent oscillations around a quadratic potential. A small deviation from a quadratic potential would not affect the physical 
outcome drastically, see~\cite{Allahverdi:2011aj}.  There are exceptions, when the scalar condensate oscillates around a displaced minimum, in which case 
there would be a kinematical blocking due to finite VEV effect, which might hamper preheating~\cite{Kofman:1997yn}, and reheating both~\cite{Allahverdi:2005mz}.}.

Let us denote $\chi$ as an ALP with a mass $m_\chi$, which couples to the $\phi$ field with derivative interactions,
and $f_{\chi}$ is the axion decay constant.
Here we will consider two scenarios:

\begin{itemize}
\item
when the axion decay constant is $f_\chi \sim \Phi_{\rm I} \sim M_P$.

\item
when the axion decay constant is $f_\chi < \Phi_{\rm I}\sim M_P$,

\end{itemize}
where we have assumed that the condensate 
is oscillating with an initial amplitude $\Phi_{\rm I} \sim {\cal O}(M_P)$~\footnote{Any light
condensate, lighter than the Hubble expansion rate during inflation can obtain large VEVs during inflation and
act as a condensate. In principle, we will be able to relax this condition also. As we shall see what matters is the 
ratio: $f_\chi/\Phi_{\rm I}$, where $\Phi_{\rm I}$ is the initial amplitude of coherent oscillations.}.
Assuming that $\phi(t)$ is a homogeneous condensate, its equation of motion in a Friedmann-Robertson-Walker (FRW) 
background is determined by:
\begin{eqnarray}
\label{eq:phieom}
\ddot{\phi} + 3 H \dot{\phi} +  m_\phi^2 \phi = 0 \,,
\end{eqnarray}
where 'dot' denotes derivative w.r.t. physical time, $t$.
A typical condensate can in principle attain a large amplitude 
during its coherent oscillations. 
The solution of the above equation is given by
\begin{eqnarray} \label{eq:phi1}
\phi (t) &\simeq& \Phi(t) \sin( m_\phi t ) \,\\
\Phi(t) & \sim &\frac{\Phi_{\rm I}}{m_\phi t}\approx \frac{ M_P}{m_\phi t}\,
\end{eqnarray}
Note that during the coherent oscillations, the average equation of state is that of matter dominated epoch, the 
Hubble expansion rate is given by: $\dot a(t)/a(t)\equiv H(t) \sim ({2}/{3t})$.
When the Hubble parameter, $ H(t) \sim m_\phi $, the condensate begins coherent oscillations.
The equation of motion for $\chi$ field is then given by~\footnote{An alternative description with canonically normalised
$\chi$ field is derived in an Appendix B.}
:
\begin{eqnarray}
\label{eq:chieom}
&& \partial^\mu \partial_\mu \chi  + \left( 3 H 
 + \frac{(\partial^0 \phi / f_\chi + \phi \partial^0 \phi / f_\chi^{2})}{( 1 + \phi / f_\chi +  \phi^{2} / 2 f_\chi^{2} )}  \right) \partial_0 \chi
 \nonumber  \\ 
&&+ \frac{m_\chi^2}{( 1 + \phi / f_\chi +  \phi^{2} / 2 f_\chi^{2} )}
 \chi = 0 
\, .
\end{eqnarray}
Expanding $\chi$ in terms of the momentum modes,
the above Eq.~(\ref{eq:chieom}) can be rewritten as:
\begin{eqnarray} \label{eq:chikeom}
\ddot{\chi}_k +  \alpha(t) \dot{\chi}_k + \omega_k^2(t) \chi_k   =0 \, ,
\end{eqnarray}
where
\begin{eqnarray} \label{eq:alpha}
\alpha(t) = 3 H(t)  + \frac{ {\dot{\phi}}/{f_\chi}  + { \phi \dot{\phi}}/{f_\chi^{2}} }{1+ {\phi}/{f_\chi} + { \phi^{2}}/{2 f_\chi^{2}} } \,,
\end{eqnarray}
and
\begin{eqnarray} \label{eq:omega}
\omega_k^2(t) = \frac{k^2}{a^2}  + \frac{m_\chi^2}{1+ {\phi}/{f_\chi} + {\phi^{2}}/{f_\chi^{2}} } \,,
\end{eqnarray}
are the damping coefficient, and the effective frequency squared, respectively. Throughout this paper we will
follow the  positive energy initial solution, i.e $\chi_{k} \simeq e^{-i \omega_k t}/\sqrt{2 \omega_k}$.   
Let us now study Eq.~(\ref{eq:chikeom}) for various scenarios.


\section{Analysis of a dimensional $5$ operator} \label{sec:dim5}

In this section, we will consider dimensional $5$ operator, i.e. $\phi \, (\partial \chi)^2 /f_\chi$, and we will assume dimensional 
$6$ operator and higher order operators are negligible. Both, $\alpha (t)$, and the $\omega_k^2(t)$, can be then approximated by
\begin{eqnarray} \label{eq:alpha_h1}
\alpha(t) & \simeq &
3 H  + \frac{ \dot{\phi}/f_\chi}{1+\phi/f_\chi } 
\nonumber \\
& \simeq &
3 H  + \frac{ m_\phi \Phi(t) \cos(m_\phi t)/f_\chi}{1+ \Phi(t) \sin(m_\phi t)/f_\chi }
\, ,
\end{eqnarray}
and
\begin{eqnarray} \label{eq:omega_h1}
\omega_k^2(t) & \simeq &
\frac{k^2}{a^2} + \frac{m_\chi^2}{1+ \phi/f_\chi }
\nonumber \\
& \simeq &
\frac{k^2}{a^2} + \frac{m_\chi^2}{1+ \Phi(t) \sin(m_\phi t)/f_\chi }
\, ,
\end{eqnarray}
respectively. We now proceed to discuss the evolution of the dynamical system for two cases: $f_\chi \sim M_P$, and  when 
and $f_\chi < M_P$.

\subsection{Axion decay constant of order Planck scale: $f_\chi \sim \Phi_{\rm I} \sim {\cal O}(M_P)$}

When  $ \Phi_{\rm I}/f_\chi \leq 1 $, where $\Phi_{\rm I} \sim M_P$, Eq.~(\ref{eq:alpha_h1}) can be approximated by:
\begin{eqnarray} \label{eq:alphahl1}
\alpha(t) &\simeq &
3 H  +  \frac{\dot{\phi}}{f_\chi} \left( 1 - \frac{ \phi }{f_\chi} \right)
\nonumber \\
&\simeq & 
3 H  +    m_\phi \cos(m_\phi t) \frac{\Phi(t) }{f_\chi} \left( 1 - \frac{\Phi(t) }{f_\chi} \sin(m_\phi t) \right)
\, .
\nonumber \\
\end{eqnarray}
Similarly Eq.~(\ref{eq:omega_h1}) can be approximated by:
\begin{eqnarray} \label{eq:omegakhl1}
\omega_k^2 &\simeq &
\frac{k^2}{a^2}+ m_\chi^2 \left( 1 - \frac{\phi}{f_\chi}  \right)
\nonumber \\
&\simeq & 
\frac{k^2}{a^2}+ m_\chi^2 \left( 1 - \frac{\Phi(t)}{f_\chi} \sin(m_\phi t)  \right)\, . 
\end{eqnarray}
Since $\Phi/f_\chi\leq 1$, the damping coefficient can be approximated by:
$\alpha  \simeq  [2 +  M_P \cos(m_\phi t)/f_\chi]/t$, during coherent oscillations of $\phi$, which is always positive. 
For the same reason, $\omega_k^2$ is always positive. Hence, one would naively think that there cannot be tachyonic instabilities, which are triggered by a negative squared frequency~\cite{Felder:2000hj,Felder:2001kt}. 

We will see later that the presence of
an oscillating damping coefficient leads to tachyonic instabilities~\footnote{This can be easily seen by means of redefinition of $\chi_k$, see for instance Eqs.(\ref{eq:transform2})-(\ref{eq:ykomega2}) for the dimensional 6 operator, a similar transformation holds for the dimensional 5 operator.}. However, as we shall see now, these tachyonic instabilties do not lead to significant enhancement in $\chi_k$ when $f_\chi \gtrsim \Phi_{\rm I}$.
Nevertheless, the evolution of the time dependent frequency, $\omega_k$, may become non-adiabatic during parts of $\phi$ oscillations, leading to {\it narrow} parametric resonances~\cite{Preheating}. We will investigate this possibility in coming section.

\subsection{Excitation of ALPs when $f_\chi \sim \Phi_{\rm I} \sim {\cal O}(M_P)$}

Let us first ignore the 
damping term and the expansion rate of the Universe by setting $\alpha =0$ and $a=1$.
Introducing the following dimensionless measure 
of time: $ m_\phi t = 2 z + \pi/2 $, one can recast Eq.~(\ref{eq:chikeom}) into the form of Mathieu equation~\cite{Mathieu}:
\begin{eqnarray} \label{eq:chikeomM1}
\chi_k'' + [A - 2 q \cos (2 z)] \chi_k = 0 \, ,
\end{eqnarray}
with
\begin{eqnarray}
 A  = \frac{4 (k^2 + m_\chi^2)}{m_\phi^2}\,,
\end{eqnarray}
and
\begin{eqnarray}
q   =  2 \frac{\Phi}{ f_\chi} \frac{m_\chi^2}{m_\phi^2} \, ,
\end{eqnarray}
where 'prime' denotes derivative w.r.t. $z$. The quanta of the $\chi$ field are produced when the evolution of the the time dependent frequency 
becomes non-adiabatic~\cite{Kofman:1994rk,Kofman:1997yn}, i.e.
\begin{eqnarray} \label{eq:adiabaticity1}
\left\vert \frac{\dot{\omega}_k}{\omega_k^2} \right\vert \simeq
\frac{m_\chi^2 \vert \dot{\phi} \vert}{2 f_\chi
\left[ k^2 + m_\chi^2 (1 - \frac{\phi}{f_\chi} )\right]^{3/2}} \gtrsim 1  \, ,
\end{eqnarray}
where $ \vert \dot{\phi} \vert \leq m_\phi \Phi$. Now we will consider two scenarios:\\

\begin{itemize}

\item{\underline{Case -1: When $k < m_\chi$:}\\
For modes with $ k < m_\chi $, the adiabaticity condition, see Eq.~(\ref{eq:adiabaticity1}), is violated when
\begin{eqnarray} \label{eq:adiabaticityklm1}
\frac{m_\chi}{m_\phi} \lesssim \frac{1}{2} \frac{\Phi}{f_\chi}   \, ,
\end{eqnarray}
which implies that $q \lesssim \frac{1}{2} (\frac{\Phi}{f_\chi})^3 \lesssim \frac{1}{2}$, in which case
the dynamical system effectively enters the narrow parametric resonance regime, 
if $A \gtrsim 1 - q$~\cite{Preheating,Kofman:1997yn}.
Eq.~(\ref{eq:adiabaticityklm1}) also defines the relation between $A$ and $q$, $A \simeq 2 (f_\chi/\Phi) q$.

For 
$f_\chi \gtrsim \Phi_{\rm I}$, 
the system never enters any tachyonic instability~\cite{Felder:2000hj,Felder:2001kt}, since $A > 2 q$ in this case. However for
$\Phi_{\rm I} \lesssim f_\chi \lesssim 1.2 \Phi_{\rm I}$,
it hits the first resonance band leading to the parametric excitation of
ALPs with masses in the range $ [4+2 \frac{\Phi}{f_\chi}]^{-1/2} m_\phi \lesssim m_\chi \lesssim \frac{\Phi}{2 f_\chi} m_\phi$ and 
momenta $ k < m_\chi$. This occurs twice per each oscillation of $\phi$ for a time interval $\Delta t \lesssim m_\phi^{-1}$, each.

During these time periods, $\chi_k$ grows as $\chi_k\approx \exp(\mu m_\phi t)$, where $\mu$ being the characteristic 
{\it Floquet} exponent~\cite{Mathieu}, which for the first resonance band is given by $\mu = \frac{1}{2} \sqrt{ A^2 - (q-1)^2}$. For the case at hand,
$\mu \lesssim \sqrt{3}/4$.

Hence,
one can estimate the maximum enhancement in the energy density of a particular momentum mode, 
$\rho_k \propto \vert \chi_k\vert^2$ (see Eq.~\ref{eq:rhok})~\footnote{See the definitions of $\rho_k(t)$, and the energy density of $\chi$ quanta, $\rho_{\chi}(t)$, in appendix \ref{app:nk}.},
during each oscillation of $\phi$ by $\delta \rho_k \sim \exp(4 \mu m_\phi \Delta t) \sim \exp(\sqrt{3})$, which means that the energy density of the modes defined by the mass and momentum range above increases at most by a factor of 6 per each oscillation of $\phi$.}\\


\item{\underline{Case -2: When $ k > m_\chi $:}\\
For modes with $ k > m_\chi $, the adiabaticity condition is violated, i.e. $\dot\omega_{k}/\omega_k^2\geq 1$, when
\begin{eqnarray} \label{eq:adiabaticitykgm1}
\frac{k}{m_\phi} \lesssim \left[ \frac{1}{2} \frac{\Phi}{f_\chi}  \frac{m_\chi^2}{m_\phi^2}   \right]^{1/3} \, ,
\end{eqnarray}
which implies that $\frac{m_\chi}{m_\phi} \lesssim \frac{\Phi}{2 f_\chi}$. Therefore, modes with 
momenta $\frac{m_\chi}{m_\phi} \lesssim \frac{k}{m_\phi} \lesssim  \frac{\Phi}{2 f_\chi} $ can be excited. Here again,
$q \lesssim \frac{1}{2} (\frac{\Phi}{f_\chi})^3 \lesssim \frac{1}{2}$, and the allowed parameter space is in the range: 
$2 (f_\chi/\Phi) q \lesssim A \lesssim (2 q)^{2/3}$. Hence, $A > 2 q$, and the system never hits the tachyonic instability of the Mathieu 
equation~\cite{Felder:2000hj,Felder:2001kt}.

Again for 
$\Phi_{\rm I} \lesssim f_\chi \lesssim 1.2 \Phi_{\rm I}$, the dynamical system hits the first resonance band twice per each oscillation of $\phi$ for a time period 
$\Delta t \lesssim m_\phi^{-1}$ each. The characteristic exponent is also $\lesssim \sqrt{3}/4$. Hence, one would expect that the energy density
of the modes defined by the mass and momentum range above to increase at most by a factor of 6 per each oscillation of $\phi$.}\\

\item{\underline{\bf By taking into account of expansion:}\\
Let us now include the effect of expansion. In this case the amplitude of both $\phi$ and $\chi_k$ oscillations die out with time. In fact, $\Phi$ drops to roughly $0.1$ of its initial value after the fist oscillation. Consequently, $q$ drops by a factor of 1000 and $A$ drops by at least a factor of 100.
As a result, the system is pushed out of the instability band, and the real part of characteristic exponent becomes 0, even before the end of the first oscillation of 
$\phi$. This render the non-perturbative production of ALPs in this scenario utterly inefficient. }

\end{itemize}

To conclude this section, one can safely conclude that for
 $f_\chi \gtrsim \Phi_{\rm I}$,
there is no significant amplification of $\chi_k$. 
In other words, there is no non-perturbative particle production of $\chi$ quanta in this case. The decay from $\phi$ condensate to $\chi$ quanta 
will be primarily perturbative, and it is given by:
\begin{eqnarray} \label{eq:gamma_perturbative}
\Gamma(\phi \rightarrow \chi \chi) = \frac{1}{64 \pi} \frac{m_\phi^3}{f_\chi^2}\,,
\end{eqnarray}
which is typical of any slow decay of an inflaton/moduli via higher dimensional operator.

\subsection{Axion decay constant smaller than Planck scale: $f_\chi < \Phi_{\rm I} \sim M_P$: Ghost like ALPs draining energy}

\begin{figure}[ht]
\begin{center}
$
\begin{array}{cc}
\includegraphics[width=8.0cm]{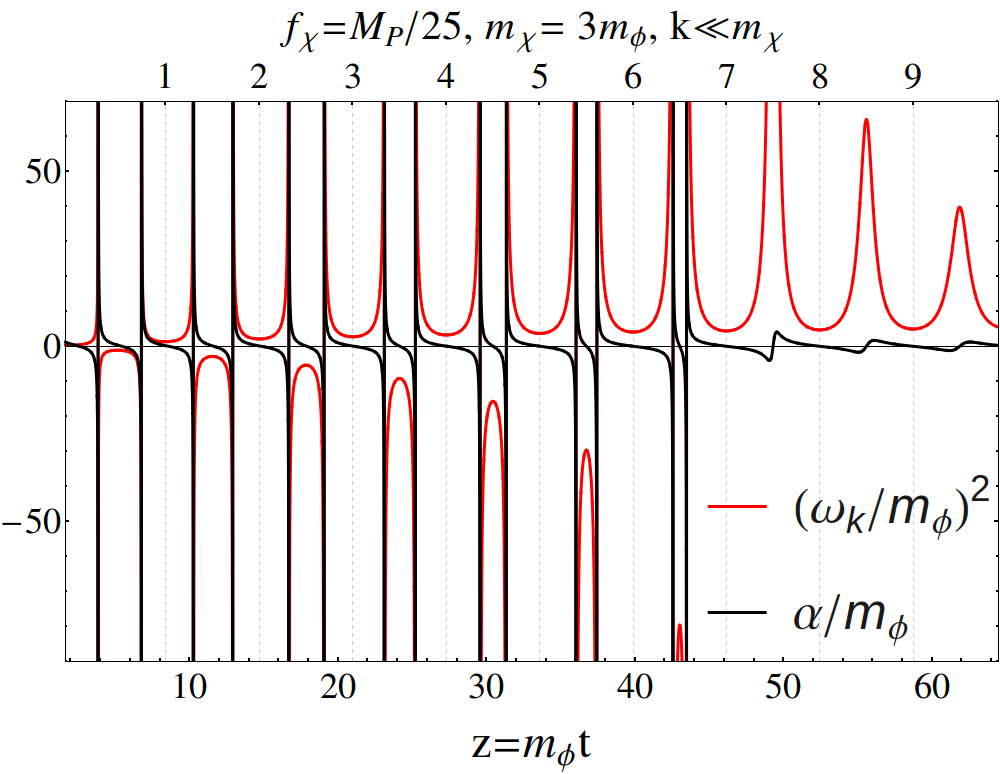}
\\
\includegraphics[width=8.0cm]{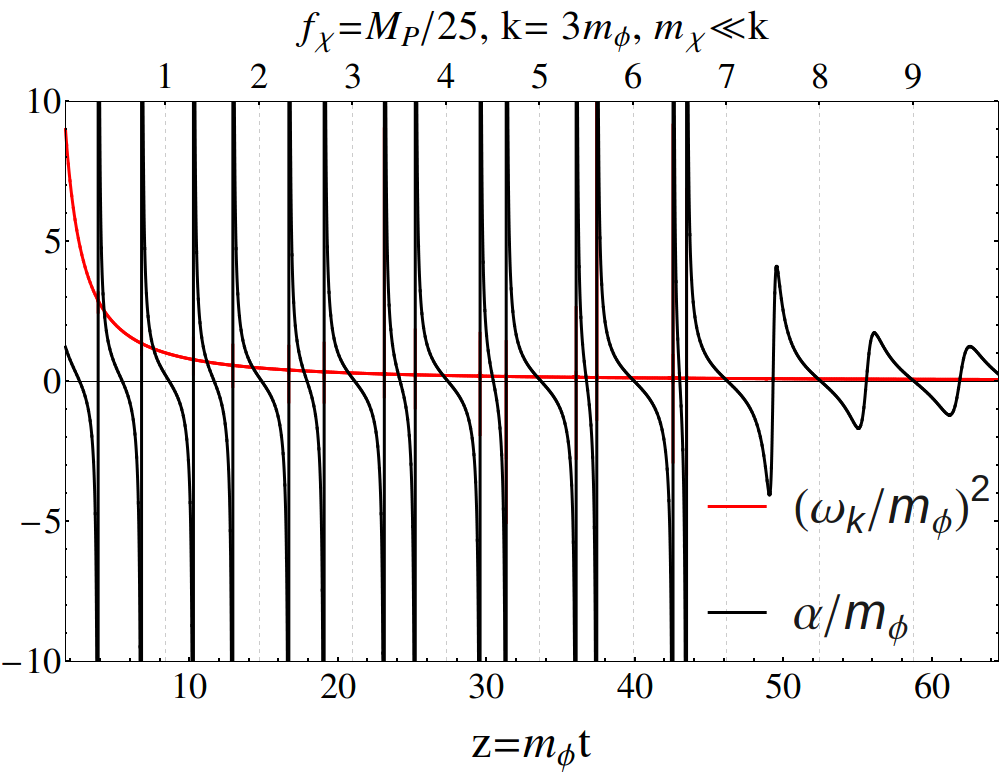}
\end{array}
$
\end{center}
\caption{The above figure shows the time evolution of  $\alpha(z)/m_\phi$ (black curve) and $\omega_k^2(z)/m_\phi^2$ (red curve) for $f_\chi=M_P/25$. The evolution of 
$\alpha(z)/m_\phi$ becomes negative roughly half the time period of oscillations, and becomes singular at the times defined by Eqs.~(\ref{eq:zi}) and (\ref{eq:zj}), this results in tachyonic instabilities.
For $k<m_\chi$, the evolution of $\omega_k^2(z)/m_\phi^2$ becomes negative for less than half the time period of oscillations as long as $\Phi/f_\chi > 1$, see the upper panel, whereas for $k>m_\chi$, the evolution $\omega_k^2(z)/m_\phi^2$ becomes negative only in the vicinity of singular points, see the lower panel.
Here time is measured in units $m_\phi^{-1}$; the corresponding number of $\phi$ oscillations is shown on the upper horizontal axis.}

\label{fig:plot1}
\end{figure}

Now, let us consider the case where  $\Phi_{\rm I}/f_\chi \gg 1$, where $\Phi_{\rm I} \sim M_P$.
For convenience, let us rewrite Eq.~(\ref{eq:chikeom}) in terms of the dimensionless measure of time, $z=m_\phi t$:
\begin{eqnarray} \label{eq:chikeomz}
\chi_k'' +  \frac{\alpha(z)}{m_\phi}  \chi_k' + \frac{\omega_k^2(z)}{m_\phi^2} \chi_k   =0 \, ,
\end{eqnarray}
where
\begin{eqnarray} \label{eq:alpha_zh1}
\alpha(z) \simeq
3 H + \frac{ m_\phi ~  \Phi \cos(z)/f_\chi }
{ 1 + \Phi \sin(z) / f_\chi } \, ,
\end{eqnarray}
and
\begin{eqnarray} \label{eq:omega_zh1}
\omega_k^2(z) \simeq \frac{k^2}{a^2} + \frac{m_\chi^2}
{ 1 + \Phi \sin(z) /f_\chi  } \, .
\end{eqnarray}
Both $\alpha(z)$ and $\omega_k^2(z)$ can be negative  
leading to an exponential growth of $\chi_k$, and therefore copious particle production, see Fig.~(\ref{fig:plot1}).
In fact, both $\alpha(z)$ and $\omega_k^2(z)$ become singular, when $\phi (t)= - f_\chi$, i.e. at
\begin{eqnarray} \label{eq:zi}
z_i^{\rm (r)} &=&  \sin^{-1}({f_\chi}/{\Phi_i}) + (2{\rm r}  - 1)\pi  \nonumber \\
&\sim &  {f_\chi}/{\Phi(z_{*, \rm r})}  + (2{\rm r} - 1)\pi  \, ,
\end{eqnarray}
and
\begin{eqnarray} \label{eq:zj}
z_j^{\rm (r)} &=& - \sin^{-1}(f_\chi/\Phi_j) + 2{\rm r}  \pi  \nonumber \\
& \sim & - {f_\chi}/{\Phi(z_{*,\rm r})} + 2{\rm r}  \pi   \, ,
\end{eqnarray}
where ${\rm r}=1,2,\ldots$ denotes the ${\rm r^{th}}$ oscillation of $\phi(t)$, and
\begin{eqnarray} \label{eq:z*}
z_{*,\rm r} = z_i^{\rm (r)} + \frac{1}{2} [z_j^{\rm (r)}-z_i^{\rm (r)}]
\sim  (4{\rm r} -1) \frac{\pi}{2}\,,
\end{eqnarray}
refers to the mid of the $\rm r^{th}$ time interval ,where $\omega_k^2$ is tachyonic, see Fig.~(\ref{fig:plot1})~\footnote{Here and 
through out this paper, we fixed the initial time, i.e. the time at which the scalar condensate starts oscillating, to $\pi/2$.}.

The number of singular points $\alpha(z)$ and $\omega_k^2(z)$ pass through depends on the ratio of $f_\chi/\Phi_{\rm I}$,
i.e. smaller this ratio is, the more singular points $\alpha(z)$ and $\omega_k^2(z)$ will have. 
Between the singular points $z_i^{\rm (r)}$ and $z_j^{\rm (r)}$, $\phi (t)< -f_\chi$, which results in a {\it ghost-like} behaviour of the $\chi$ field, as the kinetic term 
becomes negative between these singular points. This can be easily seen from the $\chi$ effective Lagrangian, which in this case is given by: 
$${\cal L}_\chi \simeq \frac{1}{2}( 1+\phi (t)/f_\chi ) (\partial \chi)^2 - \frac{1}{2} m_\chi^2 \chi^2.$$

In fact, the singular points arise when the kinetic term changes form positive to negative, and vice versa. At the singular points the coefficient of the kinetic term becomes zero. As a consequence, $(\partial \chi)^2$, and hence $\chi^2$ become infinite. Moreover, when $\phi < - f_\chi$, $\omega_k^2< 0$ become tachyonic, see Eq.~(\ref{eq:omega_zh1}), leading to an exponential production of ALPs with mass $m_\chi \gg m_\phi$ and momenta $ k \lesssim m_\chi$. This can be understood from a very rough solution,  $\chi_k \propto \exp(\vert \omega_k \vert/m_\phi ~ \Delta z)$, during the tachyonic phase. Also, the parametric excitation of ALP is possible as the evolution of $\omega_k(t)$ in time become non-adiabatic, i.e. 
\begin{eqnarray} \label{eq:adiabaticity_no_damping_mgk}
\left \vert \frac{\dot{\omega}_k}{\omega_k^2} \right \vert
\simeq
\frac{\Phi}{f_\chi} \frac{m_\phi}{2 m_\chi} \left \vert\frac{ \cos(z) }{(1+\Phi \sin(z)/f_\chi)^{1/2}  } \right \vert  \gtrsim 1 \, ,
\end{eqnarray}
for $ k < m_\chi/\vert 1+ \Phi \sin(z) /f_\chi\vert^{1/2}$, and 
\begin{eqnarray} \label{eq:adiabaticity_no_damping_mlk}
\left \vert \frac{\dot{\omega}_k}{\omega_k^2} \right \vert \simeq
\frac{\Phi}{f_\chi} \frac{m_\chi^2}{k^2} \frac{m_\phi}{2k}\frac{ \vert \cos(z) \vert}{(1+\Phi \sin(z)/f_\chi)^{2}} \gtrsim 1 \, ,
\end{eqnarray}
for $ k > m_\chi/\vert 1+ \Phi \sin(z) /f_\chi\vert^{1/2}$.

Nevertheless, one would anticipate that energy of the $\phi$ condensate can be drained even before the system actually hits the first singular point signalling the rise of the ghost-like behaviour of the $\chi$ field.
To see this, we will 
examine the behaviour of $\chi_k$ in the vicinity of the first singular point,
located at $z \simeq \pi + f_\chi/\Phi$.
Expanding $\alpha$ and $ \omega_k^2$ to first order around that point, Eq.~(\ref{eq:chikeomz}) becomes:
\begin{eqnarray} \label{eq:chik_singular2}
\chi_k'' - \frac{\Phi/f_\chi}{1+ \frac{\Phi}{f_\chi} (\pi - z) } \chi_k'
+ \left(\frac{k^2}{m_\phi^2}
+\frac{m_\chi^2/m_\phi^2}{1+ \frac{\Phi}{f_\chi} (\pi - z) } \right) \chi_k =0 \, ,
\nonumber \\
\end{eqnarray}
where for simplicity the effect of expansion was ignored.
Eq.~(\ref{eq:chik_singular2}) admits the following solution:
\begin{eqnarray} \label{eq:chik_singular2_solklm}
\chi_k &\propto&
e^{-i k z}
\left\{
C_1 ~ L_{(-l)} \left(
-2i \frac{k}{m_\phi} \frac{f_\chi}{\Phi} \left[ 1 + \frac{f_\chi}{\Phi} (\pi - z)  \right]
\right)
\right.
\nonumber \\
&&\left.
+ C_2 ~ U \left( l,1, -2i \frac{k}{m_\phi} \frac{f_\chi}{\Phi} \left[ 1 + \frac{f_\chi}{\Phi} (\pi - z)  \right] \right)
\right\}
\, ,
\nonumber \\
\end{eqnarray}
where $l = \frac{1}{2} - \frac{i}{2} \frac{m_\chi^2}{k m_\phi} \frac{f_\chi}{\Phi}$, $L_n(x)$ is the Laguerre polynomial, and $U(a,b,x)$
is the confluent hypergeometric function of the second kind (also known as the Krummer's function of the second kind)~\cite{Abramowitz}.

From Eq.~(\ref{eq:chik_singular2_solklm}), one can see that
$\chi_k$ is divergent at the first singular point, because $U(a,b,x)$ diverges logarithmically as its argument, $x$, approaches zero i.e.  as $z \rightarrow \pi + f_\chi/\Phi$. 
Hence, ALPs with arbitrary masses and momenta can be excited in the vicinity of the first singular point.

\begin{figure}[ht]
\centering
\includegraphics[width=8.0cm]{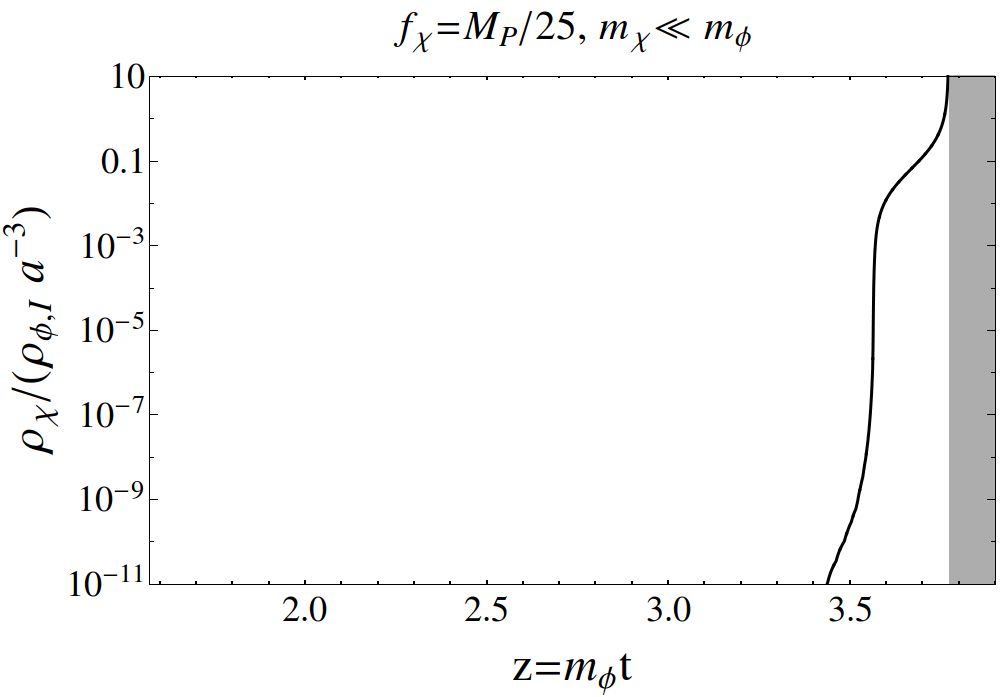}
\caption{The evolution of $\rho_\chi/(\rho_{\phi,I} a^{-3})$ during part of the $1st$ oscillation of $\phi$ field.}
\label{fig:rhochih}
\end{figure}

In particular, for ALP with a given mass, momenta in the range: $0 \leq k \lesssim (\Phi/f_\chi)(m_\chi^2/m_\phi)$ are favoured for $\chi$ production. However,  
by getting closer to the first singular point, higher momentum modes can be excited; in principal modes with infinite momenta can be excited at the exact singular 
point. However, one does have to get much closer  to the singular point to drain the energy from the condensate. See Fig.~(\ref{fig:rhochih}), where we have shown that the energy density stored in the excitations of a light $\chi$ field, $\rho_{\chi}$, is becoming comparable to that of the condensate even before the first singular point.  
The definition of the energy density of the $\chi$ field is 
defined by Eq.~(\ref{eq:rhochi}), see appendix.

For heavy ALPs, with $m_\chi \gg m_\phi$, the $\chi$ energy density becomes comparable to the scalar condensate even earlier. This renders the 
problem non-linear even before the first singular point, whose analysis goes beyond the scope of the current paper.

%
%
%
%

The above picture will be the case for any odd-dimensional operator, such as dimensions $5,~7,~9,\cdots$.
For even dimensional operators, such as $6,~8,~10,\cdots$, the ghost problem will not arise and we will be able to discuss the ALPs dynamics safely.

We now conclude our study of dimensional $5$ operator by briefly mentioning that it is indeed possible to excite copious ALPs with a wide range of masses and momenta during the first oscillation of the $\phi$ condensate. Following the evolution through out will be a daunting task analytically, since the energy density of the ALP very soon becomes comparable to that of the energy density of the condensate, besides the existence of formidable singularities. Furthermore, ghost-like behaviour of the ALP arises due to the kinetic term oscillating to negative values. This does not happen for ALP interaction with even dimensional operators, which we will study in the next section.

\section{Copious production of ALPs via dimension $6$ operator} \label{sec:dim6}

Let us assume that the ALP is coupled to the condensate only via dimension $6$ operator, then
the damping coefficient, $\alpha(t)$, can be approximated by:
\begin{eqnarray} \label{eq:alpha_h2}
\alpha(t) &\simeq & 
 3 H  + \frac{ \phi \dot{\phi}/ f_\chi^{2} }{1 + \phi^{2}/2 f_\chi^{2} }
\nonumber \\
&\simeq &
3 H  + \frac{ m_\phi \Phi^2 \sin(2 m_\phi t)/ 2 f_\chi^{2} }{1 + \Phi^{2} \sin^2(m_\phi t)/2 f_\chi^{2} }
\,,
\end{eqnarray}
similarly, $\omega_k^2(t)$ can be approximated by:
\begin{eqnarray} \label{eq:omega_h2}
\omega_k^2(t) &\simeq &
\frac{k^2}{a^2}  + \frac{m_\chi^2}{1 + \phi^{2}/2 f_\chi^{2} } 
\nonumber \\
&\simeq &
\frac{k^2}{a^2}  + \frac{m_\chi^2}{ 1 +  \Phi^{2} \sin^2(m_\phi t)/2 f_\chi^{2} } 
\, .
\end{eqnarray}
Figs.~(\ref{fig:omega_alpha_1}) and (\ref{fig:omega_alpha_2}) show plots of $\alpha/m_\phi$ (black curve) and $\omega_k/m_\phi$ (blue curve for $m_\chi \ll k$ and red curve for $k \ll m_\chi$)  for $f_\chi \simeq 1.1 \Phi_{\rm I}$ and $f_\chi = \Phi_{\rm I}/25$, respectively.
We now proceed to discuss the evolution of the dynamical system for two cases: $f_\chi \gtrsim \Phi_{\rm I} $ and $f_\chi < \Phi_{\rm I} $.

\subsection{Axion decay constant of order Planck scale: $f_\chi \sim \Phi_{\rm I} \sim {\cal O}(M_P)  $}

\begin{figure}[ht]
\begin{center}
\includegraphics[width=8.0cm]{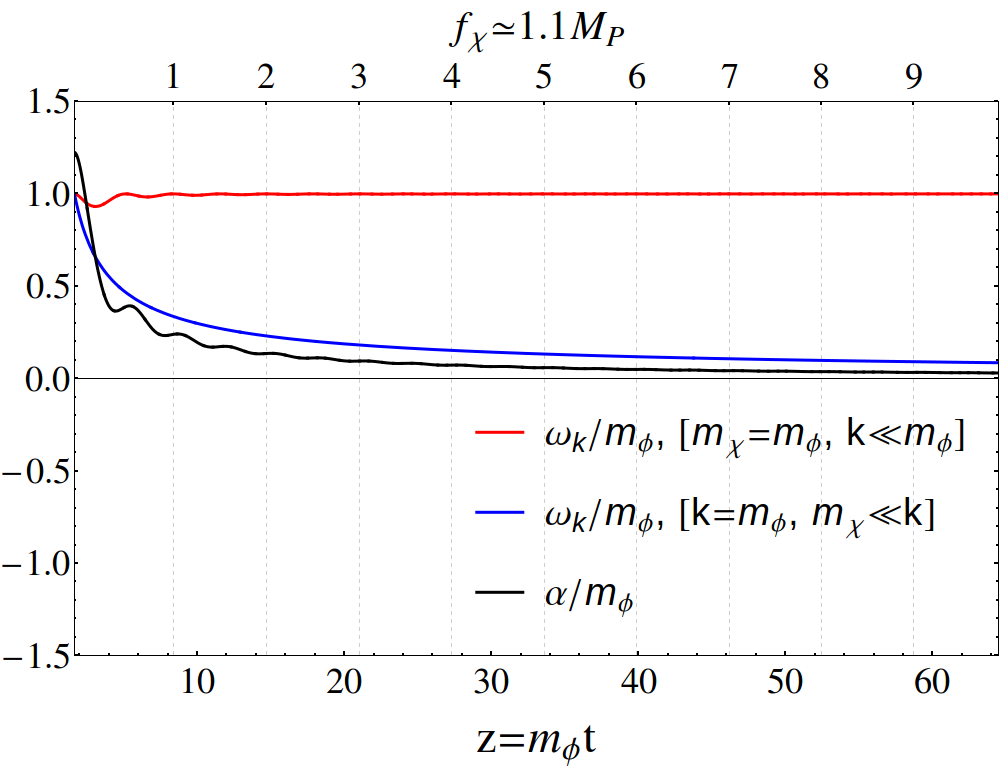}
\end{center}
\caption{The above figures show the time evolution of $\alpha/m_\phi $ (black curve), and $\omega_k/m_\phi $ for $ k\gg m_\chi$ (blue curve) and $k \ll m_\chi$ 
(red curve) for $f_\chi \simeq 1.1 M_P$. For $m_\chi \ll k$, $\omega_k$
does not oscillate and redshifts with the expansion of the Universe, whereas for
$k \ll m_\chi$, $ \omega_k$ oscillates with double the frequency of $\phi$ field, and moves toward its maximum value, $m_\chi$, as the amplitude of the oscillations 
dies out with cosmic  expansion.
Here time is measured in units $m_\phi^{-1}$; the corresponding number of $\phi$ oscillations is shown on the upper horizontal axis.}
\label{fig:omega_alpha_1}
\end{figure}

When 
$\Phi_{\rm I}/f_\chi \lesssim 1 $, where $\Phi_{\rm I} \sim M_P$, the damping coefficient can be approximated by:
\begin{eqnarray} \label{eq:alphahl1}
\alpha(t) &\simeq &
  3 H  +  \frac{\phi \dot{\phi}}{f_\chi^{2}} 
  \left( 1 - \frac{\phi^{2}}{2 f_\chi^{2}} \right)
\nonumber \\
&\simeq & 
3 H + m_{\phi} \frac{\Phi^2}{2 f_\chi^{2}} \sin(2 m_\phi t) \left( 1- \frac{\Phi^{2}}{2 f_\chi^{2}} \sin^2(m_\phi t) \right)
\, .\nonumber \\
\end{eqnarray}
Similarly, $\omega_k^2(t)$ can be approximated by:
\begin{eqnarray} \label{eq:omegakhl1}
\omega_k^2 &\simeq&
 \frac{k^2}{a^2}+ m_\chi^2 \left( 1 - \frac{\phi^{2}}{2 f_\chi^{2}} \right)
\nonumber \\
&\simeq & 
\frac{k^2}{a^2}+m_\chi^2 \left( 1- \frac{\Phi^{2}}{2 f_\chi^{2}} \sin^2(m_\phi t) \right)
\, .
\end{eqnarray}
For $ \Phi/f_\chi \lesssim 1$,
$\alpha  \simeq  2 t^{-1} +  \Phi_{\rm I}^2(2  m_\phi t^2 f_\chi^2)^{-1} \sin(2 m_\phi t)$ is always positive, which can be also seen from Fig.~(\ref{fig:omega_alpha_1}). 
For the same reason, $\omega_k^2$ is always positive. However, due to the presence of an oscillating damping coefficient, tachyonic instabilities appear for short time intervals while $\phi$ is oscillating. This can be easily seen from Eqs.(\ref{eq:transform2})-(\ref{eq:ykomega2}). However, as will see below, these tachyonic instabilties do not lead to significant enhancement in $\chi_k$ when $f_\chi \gtrsim \Phi_{\rm I}$.
Moreover as $\phi$ oscillates, the evolution of $\omega_k$ in time may become non-adiabatic giving rise to parametric instabilities which may lead to exponential enhancement in $\chi_k$.
We will investigate this possibility below.
%
%
%
%

In order to investigate the existence of such parametric resonance,
let us for the time being ignore the 
damping term and the momentum redshift, i.e. by setting $\alpha =0$ and $a=1$.
Introducing the dimensionless measure of time, $z = \pi/2 +  m_\phi t $,
 Eq.~(\ref{eq:chikeom}) can be put into the form of Mathieu equation~\cite{Mathieu}
\begin{eqnarray} \label{eq:chikeomM2}
\chi_k'' + [A - 2 q \cos (2 z)] \chi_k = 0 \, ,
\end{eqnarray}
where
\begin{eqnarray}
A &= &
\frac{k^2}{m_\phi^2} + \frac{m_\chi^2}{m_\phi^2} \left( 1- \frac{\Phi^2}{4 f_\chi^2} \right) \,,\\
q &=&
\frac{\Phi^{2}}{ 8 f_\chi^{2}} \frac{m_\chi^2}{m_\phi^2}  \,.
\end{eqnarray}
Again, the quanta of the field $\chi_k$ are produced when the adiabaticity condition gets violated, i.e.
\begin{eqnarray} \label{eq:adiabaticity2}
\left\vert \frac{\dot{\omega}_k}{\omega_k^2} \right\vert =
\frac{ \Phi^2}{4 f_\chi^2}
\frac{ m_\chi^2 m_\phi \vert \sin(2 m_\phi t) \vert}
{\left[ k^2 + m_\chi^2 \{ 1 - \frac{ \Phi^2}{2 f_\chi^2} \sin^2(m_\phi t) \} \right]^{3/2}} \gtrsim 1  \, . \nonumber \\
\end{eqnarray}
Let us consider two scenarios:

\begin{itemize}

\item{\underline{For modes with $ k < m_\chi $}: the adiabaticity condition is violated, when
\begin{eqnarray} \label{eq:adiabaticityklm2}
\frac{m_\chi}{m_\phi} \lesssim  \frac{ \Phi^2}{4 f_\chi^2}
\, .
\end{eqnarray}
Hence, $q = \frac{1}{128} (\frac{\Phi}{f_\chi})^6 \ll1$,
which implies that the dynamical system enters the resonance regime, if $A \gtrsim 1 - q$.
However, here $A \lesssim \frac{1}{16} (\frac{\Phi}{f_\chi})^4 < 1 - q$. 
Hence, the dynamical system never hits the resonance bands of the Mathieu equation. 
Moreover, since $A > 2 q$, the system does not enter the tachyonic instability region. 
Therefore, there is no significant non-perturbative particle production in this case.}

\item{\underline{For modes with $ k > m_\chi $}:
the adiabaticity condition is violated for
\begin{eqnarray} \label{eq:adiabaticitykgm2}
\frac{k}{m_\phi} \lesssim \left[ \frac{\Phi^2}{4 f_\chi^2} \frac{m_\chi^2}{m_\phi^2}   \right]^{1/3}
\, ,
\end{eqnarray}
which implies that $ \frac{m_\chi}{m_\phi} \lesssim \frac{\Phi^2}{4 f_\chi^2}$. As a result,
$q = \frac{1}{128} (\frac{\Phi}{f_\chi})^6 \ll1$ and $A \lesssim \frac{1}{16} (\frac{\Phi}{f_\chi})^4 < 1 - q$. Again, $A$ is never below $2 q$. 
Therefore, the dynamical system never hits the instability bands of the Mathieu equation.}

\end{itemize}

Including the expansion effect and the damping term, which is always positive in this case, will make 
things even worse, as far as non-adiabatic excitation of ALPs is concerned.
Therefore, one can safely conclude that for $f_\chi \gtrsim \Phi_{\rm I}$,
there is no significant amplification of $\chi_k$. In this case, 
the production of the ALPs will be primarily a
perturbative process.

\subsection{Axion decay constant smaller than Planck scale: $f_\chi \ll \Phi_{\rm I} \sim M_P $}

\begin{figure}[ht]
\begin{center}
\includegraphics[width=8.0cm]{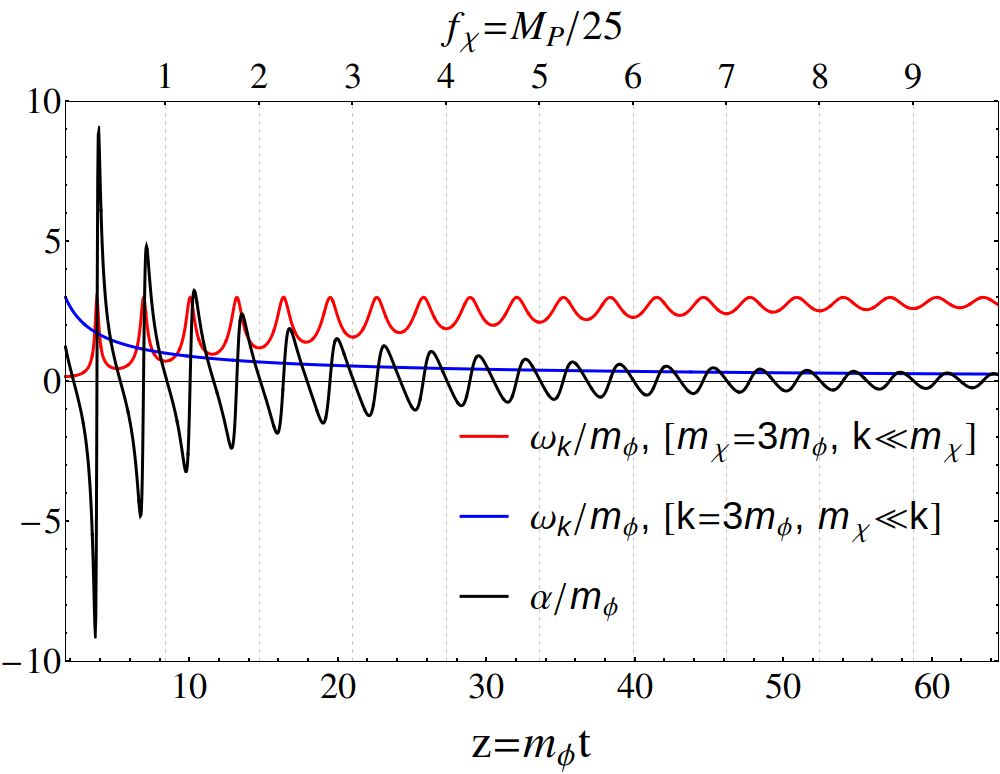}
\end{center}
\caption{Same as Fig.~(\ref{fig:omega_alpha_2}), but for $f_\chi=M_P/25$. Note that the damping coefficient, $\alpha $, now oscillates to negative values twice per each 
oscillation of $\phi$ field.}
\label{fig:omega_alpha_2}
\end{figure}

Now, let us consider the case where  $\Phi_{\rm I}/f_\chi \gg 1$, with $\Phi_{\rm I} \sim M_P$. 
In this case, the damping coefficient, $\alpha$, oscillates with a large amplitude until $\Phi$ drops roughly below $f_\chi$, see  Fig.~(\ref{fig:omega_alpha_2}).
This is actually crucial for particle creation. As we shall see, this gives rise to tachyonic instability, and parametric resonance, leading to exponential particle production. The details 
will be  discussed in the following subsections.
%
%
%
%
For convenience, let us introduce the following transformation:
\begin{eqnarray} \label{eq:transform2}
\chi_k = {\cal X}_k e^{-\frac{1}{2} \int_{z_{\!_I}}^{z} \frac{ \phi \phi' / f_\chi^2}{1+ \phi^2 /2 f_\chi^2} d\tilde{z}}  \, .
\end{eqnarray}
As a result, Eq.~(\ref{eq:chikeomz}) becomes
\begin{eqnarray} \label{eq:ykeom2}
{\cal X}_k'' +  \frac{3 H}{m_\phi} {\cal X}_k'+\frac{\tilde{\omega}_k^2}{m_\phi^2} ~ {\cal X}_k = 0 \, ,
\end{eqnarray}
with 
\begin{eqnarray} \label{eq:ykomega22}
\tilde{\omega}_k^2 & = & \frac{k^2}{a^2}
+ \frac{m_\chi^2
- \frac{m_\phi^2}{2f_\chi^2} (\phi \phi'' + \phi'^2) -\frac{3 H}{2 f_\chi^2} m_\phi\phi \phi'}{1+ \phi^2/2 f_\chi^2 }
\nonumber \\
&&+ \frac{ \frac{m_\phi^2}{4 f_\chi^4} (\phi \phi')^2}{(1+ \phi^2/2f_\chi^2)^2}  \, ,
\end{eqnarray}
being square of the frequency of ${\cal X}_k$ oscillation.

\subsubsection{{\bf Non-perturbative excitations of ALPs, ignoring cosmic expansion}}

We first start by discussing the non-perturbative excitation of $\chi$ quanta while ignoring
the effect of expansion.
In this case, $\tilde{\omega}_k^2$ becomes
\begin{eqnarray} \label{eq:ykomega2}
\tilde{\omega}_k^2 =  k^2
+ \frac{m_\chi^2
- m_\phi^2 \frac{\Phi^2}{f_\chi^2} \cos(2z)}{1+ \frac{\Phi^2}{2 f_\chi^2} \sin^2(z)}
+ \frac{ m_\phi^2 \frac{\Phi^4}{16 f_\chi^4} \sin^2(2z)}{(1+ \frac{\Phi^2}{2f_\chi^2} \sin^2(z))^2}  \, . \nonumber \\
\end{eqnarray}
Moreover, the exponential part of Eq.~(\ref{eq:transform2}) can now be easily evaluated to hold
\begin{eqnarray} \label{eq:oscill2}
\exp \left[{-\frac{1}{2} \int_{z_{\!_I}}^{z}
\frac{\phi \phi'^2/2f_\chi^2}{1+ \phi^2/2f_\chi^2 } d\tilde{z}} \right]
=
\frac{\sqrt{1 + \frac{\Phi^2}{2f_\chi^2}}}{\sqrt{1 + \frac{\Phi^2}{2f_\chi^2} [1 - \cos(2z)]}}
 \, .
 \nonumber \\
\end{eqnarray}
This term clearly exhibits an oscillatory behaviour with period $\pi$, measured in units of  $m_\phi^{-1}$. Hence, it does not lead to any enhancement of $\chi_k$.
Moreover upon the inclusion of the expansion effect, $\Phi$ dies out with time and hence such oscillatory behaviour goes away in a few oscillation of $\phi$.
It is worth noting that the above term is independent of both
$m_\chi$ and $k$.
We now proceed to discuss non-perturbative production mechanisms for two mass regimes: (1) light ALPs with mass $ m_\chi \ll m_\phi$ and (2) heavy ALPs with $ m_\chi \gtrsim m_\phi$.\\

\paragraph{\centering { {\bf Narrow resonance production for light ALPs, with $ m_\chi \ll m_\phi$} }\\}
\noindent

\begin{figure}[ht]
\centering
\includegraphics[width=8.0cm]{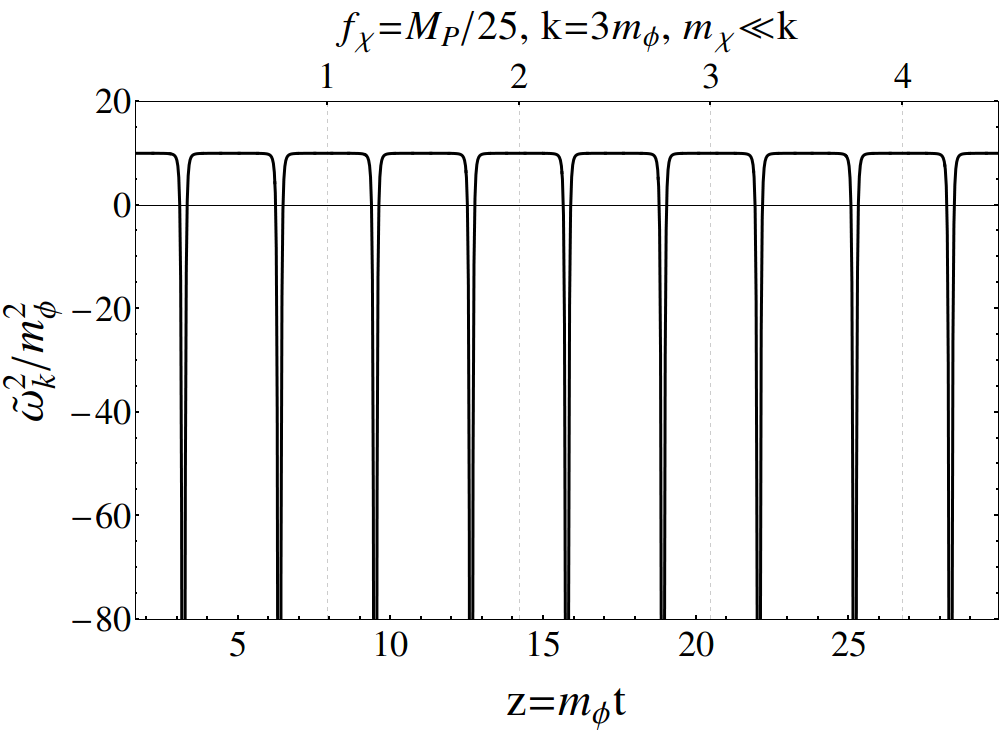}
\caption{The time evolution of $\tilde{\omega}_k^2/m_\phi^2$ for $k=3 m_\phi$ and $ m_\chi \ll k$, where the effect of expansion is ignored. The
$\tilde{\omega}_k^2$ becomes negative for two short periods of time during each oscillation of $\phi$. Here time is measured in units $m_\phi^{-1}$; the corresponding number of $\phi$ oscillations is shown on the upper horizontal axis.}
\label{fig:omega_k_no}
\end{figure}

We can see from Eq.~(\ref{eq:ykomega2}), see also Fig.~(\ref{fig:omega_k_no}), that for modes with
$ k \lesssim m_\phi \Phi/\sqrt{2} f_\chi$,
$\tilde{\omega}_k$
becomes tachyonic during short time intervals, around $z = n \pi, ~ n=0,1,2, \cdots$.
More precisely,
$\tilde{\omega}_k^2$
becomes negative during the time intervals, $\Delta z = z_+ -z_-$,
where
$$z_{\pm} \simeq n \pi \pm \left[\frac{m_\phi}{k} \frac{\sqrt{2} f_\chi}{ \Phi} - \frac{2 f_\chi^2}{\Phi^2} \right]^{1/2} $$
for $k>m_\phi$, whereas for
$k \ll m_\phi$, we have
$$z_{\pm} \simeq n \pi \pm [{f_\chi}/{\Phi}]^{1/2}.$$
This leads to tachyonic excitation of $\chi$ modes with momenta in the range
$0 \leq k \lesssim m_\phi \Phi/\sqrt{2} f_\chi$.
The upper bound on
$k$
can be easily seen from Eq.~(\ref{eq:ykomega2}), because
$\tilde{\omega}_k^2$
is positive for higher momenta.
%
%
%
On the other hand, from Fig.~(\ref{fig:omega_k_no}), one can see that
$\tilde{\omega}_k^2$
sharply changes around $z = n \pi$, before it becomes negative.
Hence, one would expect violation of the adiabaticity condition around the time intervals where $\tilde{\omega}_k^2<0$. Indeed, this is the case provided that
$k \lesssim m_\phi \Phi/\sqrt{2} f_\chi$.
More percisely, the adiabaticity gets violated during the time intervals,
$\Delta \tilde{z} =\tilde{z}_+ - \tilde{z}_-$,
where
$$\tilde{z}_{\pm} \simeq n \pi \pm \left[\frac{m_\phi^3}{k^3} \frac{4 f_\chi^2}{ \Phi^2} \right]^{1/5} \,$$
for $k>m_\phi$,
while for
$k \ll m_\phi$,
$$\tilde{z}_{\pm} \simeq n \pi \pm [{4 f_\chi^2}/{ \Phi^2}]^{1/5} \, .$$
This leads to the {\it parametric} excitation of $\chi$ quanta with momenta,
$ k \lesssim  m_\phi \Phi/\sqrt{2} f_\chi$.

Recall that $\tilde{\omega}_k^2$ becomes tachyonic during the time intervals $\Delta z= z_+ - z_-$,
around $z=n \pi$, where $z_{\pm} \simeq n \pi \pm [\frac{m_\phi}{k} \frac{\sqrt{2} f_\chi}{\Phi} - \frac{2 f_\chi^2}{ \Phi^2}]^{1/2} $ for $k>m_\phi$,
and $z_{\pm} \simeq n \pi \pm [{f_\chi}/{\Phi}]^{1/2}$ for $k \ll m_\phi$.
Hence, the parametric enhancement of
 $\chi_k$ takes place only for very short time periods around the tachyonic dimples in 
$\tilde{\omega}_k^2$, see Fig.~(\ref{fig:omega_k_no}).
In other words, both production mechanisms go side by side.

In fact, both the instabilities discussed above are caused by the oscillating damping coefficient, Eq.~(\ref{eq:alpha_h2}), that oscillates with a large amplitude, see Fig.~(\ref{fig:omega_alpha_2}),  which gives rise to deep tachyonic dimples in $\tilde{\omega}_k^2$. Moreover, parametric instabilities occurs on the sides of those tachyonic dimples before $\tilde{\omega}_k^2$ becomes negative. These instabilities lead to exponential production of ALPs.

It is worth noting here that even when $\Phi_{\rm I} \lesssim f_\chi$, these tachyonic dimples will be present when $k, m_\chi < m_\phi \Phi/\sqrt{2} M_P$, albeit being way shallower. Hence, one would expect that all modes with $ k,m_\chi \ll m_\phi$, can be excited. However, the production of such modes is insignificant as the enhancement in $\chi_k$ after each tachyonic phase, which is roughly given by $ \delta \chi_k \leq exp\{ \Phi/(2\sqrt{2}f_\chi) ~ \Delta z\} $, is at most a factor of 2 per each oscillation of $\phi$; consequently, $\rho_k$ gets enhanced by at most a factor of 4 per each oscillation of $\phi$. For the dimensional 5 operator discussed earlier, $\rho_k$ gets enhanced by roughly the same factor per each oscillation of $\phi$. However, this enhancement will go away upon the inclusion of the expansion effect.
%
%
%
\begin{figure}[ht]
\centering
\includegraphics[width=8.0cm]{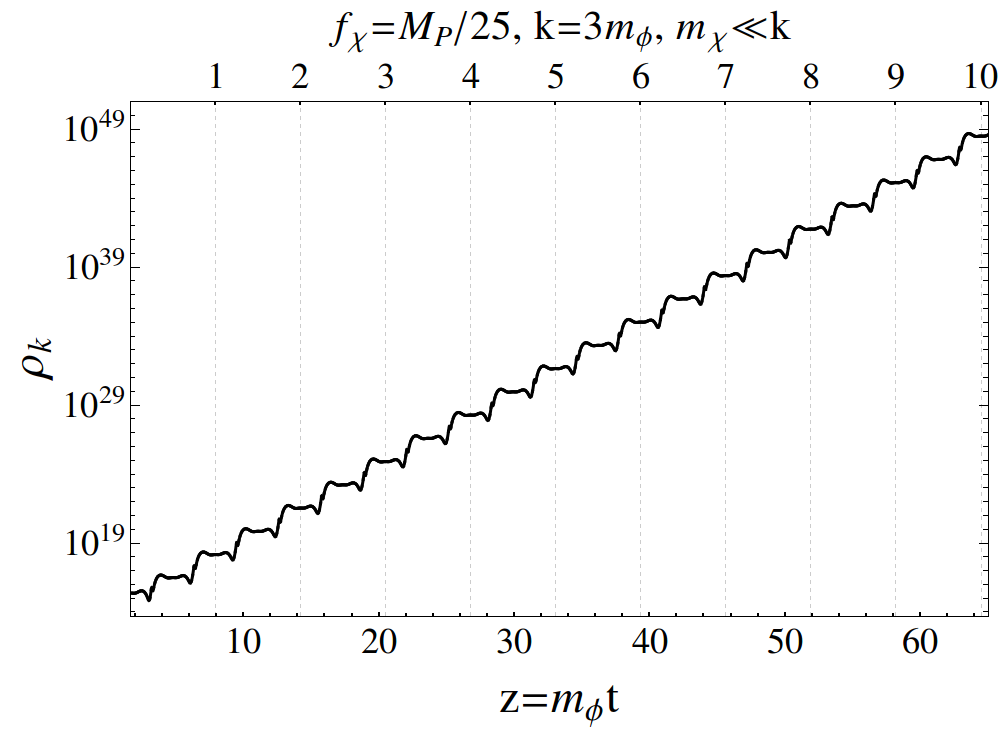}
\caption{The above figure shows time evolution of  $\rho_k$ for $k=3 m_\phi$ and $ m_\chi \ll k$, where the effect of expansion is neglected. Here time is measured in units $m_\phi^{-1}$; the corresponding number of $\phi$ oscillations is shown on the upper horizontal axis.}
\label{fig:rhok_k_no}
\end{figure}
%
%

In Fig.~(\ref{fig:rhok_k_no}), we have shown the time evolution of $\rho_k$, defined by Eq.~(\ref{eq:rhok}), for $f_\chi = M_P/25$, $m_\chi \ll m_\phi$, and $k= 3m_\phi$, while ignoring the effect of expansion.
One can see that $\rho_k$
exhibits an oscillatory behaviour, except for short time intervals, where the aforementioned instabilities resides. During these time intervals, $\rho_k$ increases rapidly to stabilize at a larger value after each period of instability.

For the parameters given above, $\rho_k$ grows roughly by a factor of 2000 per each oscillation of $\phi$, which is a very high production rate. \\

\paragraph{\centering {\bf Non-perturbative production of heavy ALPs, with $ m_\chi \gtrsim m_\phi$}\\}
\noindent

The perturbative production of ALPs with $m_\chi> m_\phi/2 $ is kinematically forbidden. However,
the production of such heavy particles can still take place during the coherent oscillations of $\phi$ condensate 
via non-perturbative effects.

\begin{figure}[ht]
\begin{center}
\includegraphics[width=8.0cm]{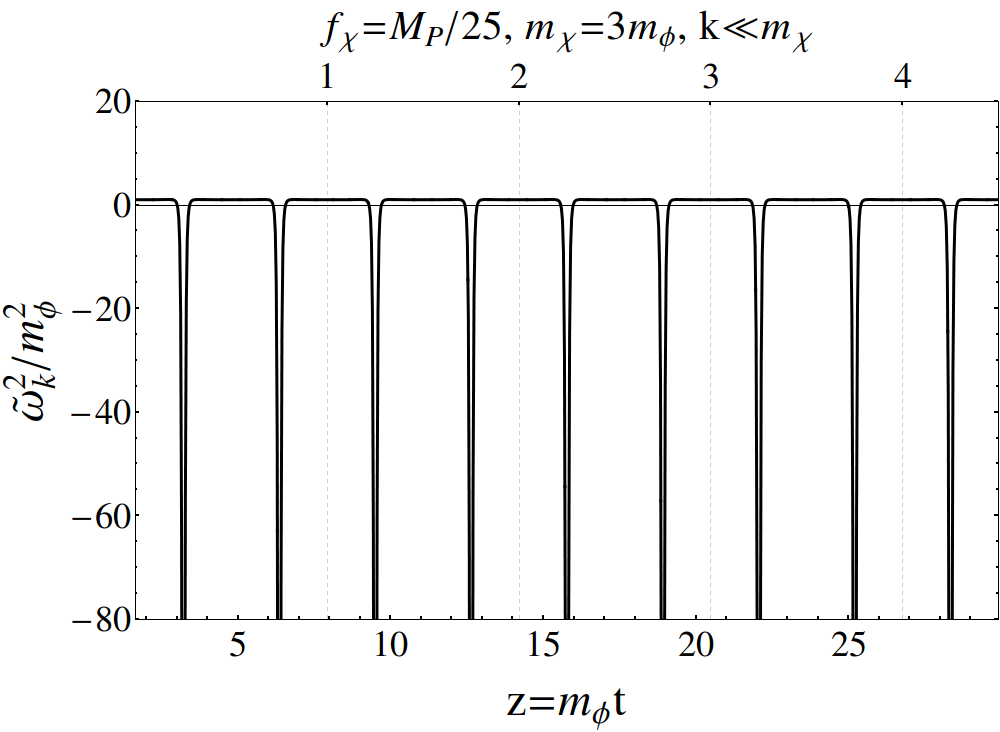}
\end{center}
\caption{Same as Fig.~(\ref{fig:omega_k_no}) but for $m_\chi =3 m_\phi$ and $k \ll m_\phi$, without taking into account of expansion.}
\label{fig:omega_m_no}
\end{figure}

Similar to the case of light ALPs, $\tilde{\omega}_k^2$ becomes tachyonic for short time periods, around 
$z = n \pi, ~ n=0,1,2, \cdots$, provided that
$ k, m_\chi \lesssim m_\phi \Phi/\sqrt{2} f_\chi$.
This occurs twice per each oscillation of $\phi$, see Eq.~(\ref{eq:ykomega2}) and Fig.~(\ref{fig:omega_m_no}).
More precisely, $\tilde{\omega}_k^2 < 0$  during the time intervals,
$\Delta z = z_+ -z_-$ around $z = n \pi$, where
$$z_{\pm} = n \pi \pm \left[\frac{m_\phi^2}{m_\chi^2} - \frac{2 f_\chi^2}{\Phi^2} \right]^{1/2}$$
for $k <m_\chi$, and 
$$z_{\pm} = n \pi \pm \left[\frac{m_\phi}{k} \frac{\sqrt{2} f_\chi}{\Phi} - \frac{2 f_\chi^2}{ \Phi^2} \right]^{1/2} $$
for $k >m_\chi$.
During these time periods, {\it tachyoinc} excitation of
$\chi$ quanta with momenta in the range $0 \leq k \lesssim \Phi/\sqrt{2} f_\chi$ takes place.
%
%
%
%
%
\begin{figure}[ht]
\centering
\includegraphics[width=8.0cm]{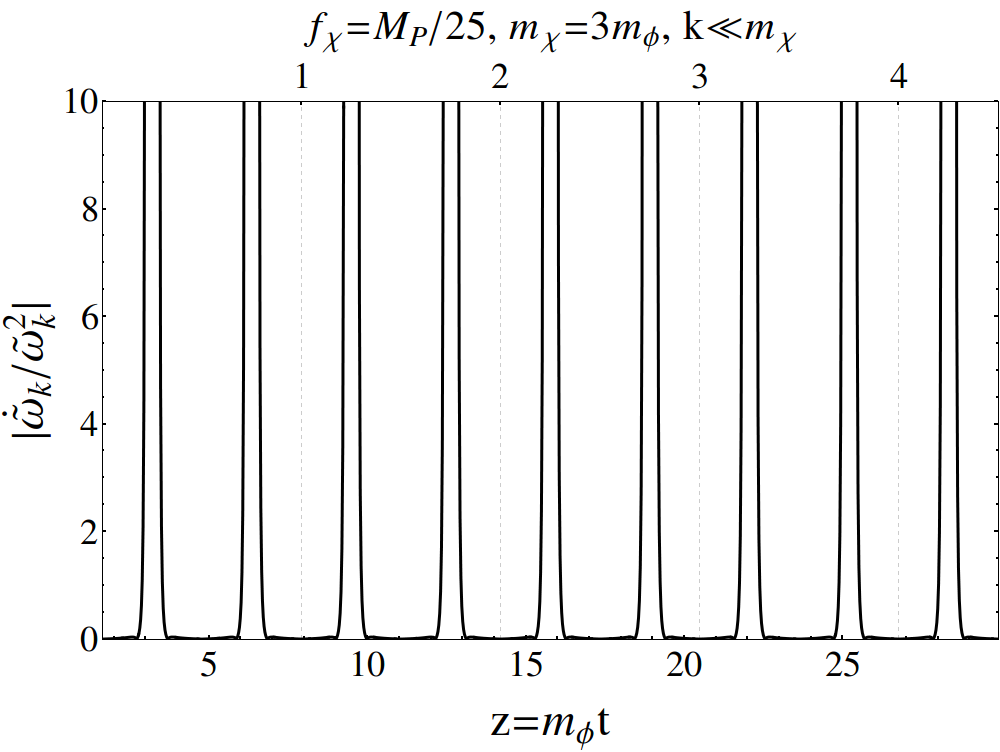}
\caption{The evolution of $\vert \dot{\tilde{\omega}}_k/\tilde{\omega}_k^2 \vert$ in time for $m_\chi =3 m_\phi$ and $k \ll m_\phi$, where the effect of expansion is ignored. One can see easily that the adiabaticity gets violated around the intervals where $\tilde{\omega}_k^2$ becomes tachyonic. This takes place twice per each oscillation of $\phi$. 
Here the time is measured in units $m_\phi^{-1}$; the corresponding number of $\phi$ oscillations is shown on the upper horizontal axis.}
\label{fig:adiab_m_no}
\end{figure}
%
%
%
%
\begin{figure}[ht]
\centering
\includegraphics[width=8.0cm]{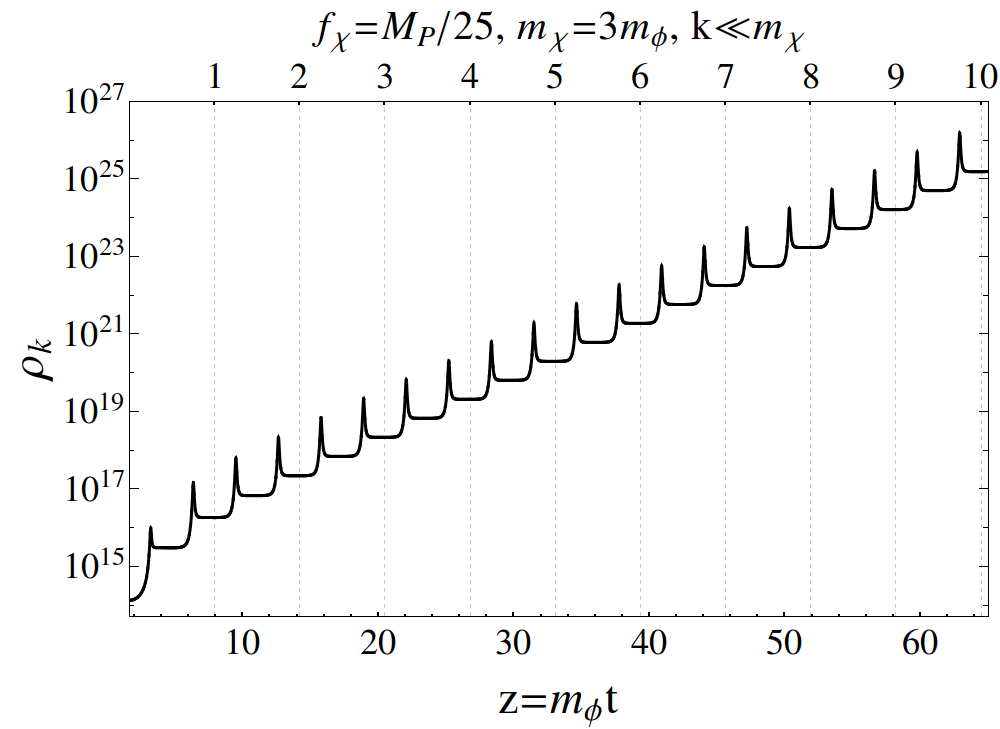}
\caption{Same as Fig.~(\ref{fig:rhok_k_no}), but for $m_\chi =3 m_\phi$ and $k \ll m_\phi$, without taking into account of expansion.}
\label{fig:rhok_m_no}
\end{figure}
%
%

In Fig.~(\ref{fig:omega_m_no}), we have shown the time evolution of $\tilde{\omega}_k^2$ for $f_\chi =M_P/25$, $m_\chi=3 m_\phi$ and $ k\ll m_\phi$ while ignoring the effect of expansion.
Similar to the case $m_\chi < k$, one can see that for $ m_\chi > k$, $\tilde{\omega}_k^2$ changes sharply around the time intervals of tachyonic instabilities.
Again one would expect violation of the adiabaticity condition during the time intervals where $\tilde{\omega}_k^2$ changes sharply. Indeed this is the case; the evolution of $\tilde{\omega}_k$
becomes non-adiabatic for short time intervals, around the tachyonic dimples residing in the vicinity of $z = n \pi$,
see Fig.~(\ref{fig:adiab_m_no}).
This leads to the parametric excitation of heavy $\chi$ quanta with momenta $k<m_\chi$. Moreover, analogous to the case of light ALPs, tachyonic and parametric production of particles with $m_\chi<k \lesssim m_\phi \Phi/\sqrt{2} f_\chi$ is also possible.

Therefore, during the short time intervals of parametric and tachyonic instabilities, heavy ALPs with momenta in the range $ 0 \leq k \lesssim m_\phi \Phi/\sqrt{2} f_\chi$ can be excited. Again, very similar to the case of light ALPs,
one can easily see from Eq.~(\ref{eq:alpha_h2}) and Fig.~(\ref{fig:omega_alpha_2}) that the instabilities discussed above are induced by the oscillating damping coefficient.

In Fig(\ref{fig:rhok_m_no}), we show the time evolution of $\rho_k$ for $m_\chi = 3 m_\phi$, $k \ll m_\chi$ and $f_\chi = \Phi_{\rm I}/25$ while ignoring the effect of expansion. One can easily see that $\rho_k$ gets largely enhanced at each interval of instability which happens twice per each oscillation of $\phi$. However comparing this to Fig(\ref{fig:rhok_k_no}), one can see clearly that the production of ALPs with momenta, $k>m_\chi,m_\phi$ is way more efficient than those with momenta, $k <m_\chi$.
%
%
\begin{figure}[ht]
\centering
\includegraphics[width=8.0cm]{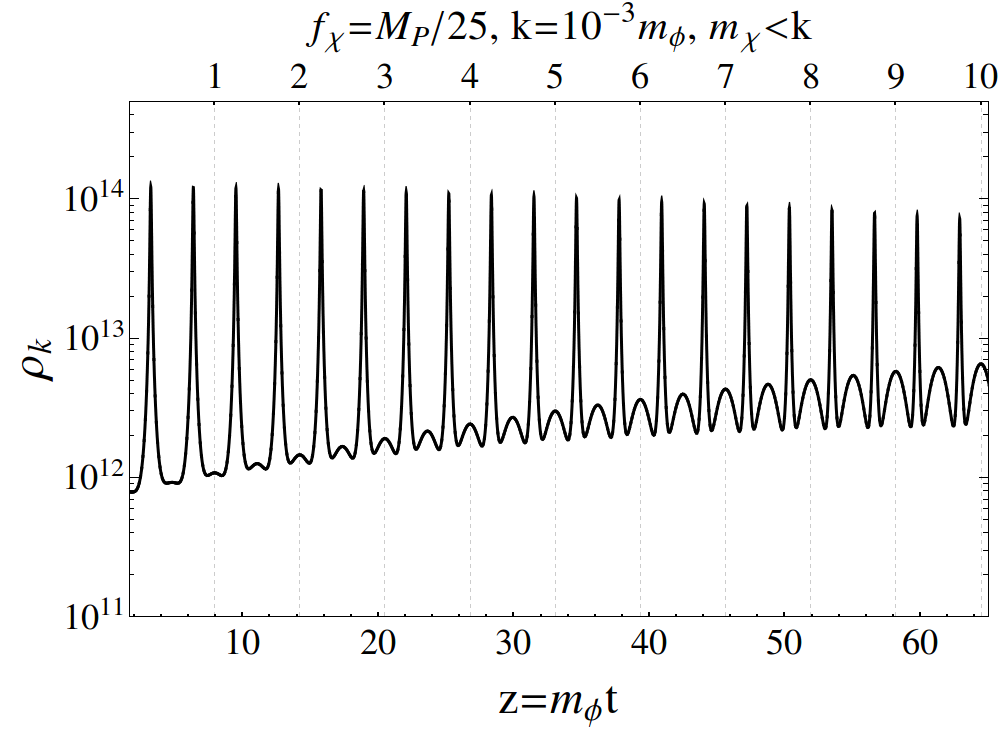}
\caption{The evolution of $\rho_k$ in time for $f_\chi = M_P/25$ and $m_\chi < k \ll m_\phi$ where the effect of expansion is included. Here the time is measured in units $m_\phi^{-1}$; the corresponding number of $\phi$ oscillations is shown on the upper horizontal axis.}
\label{fig:rhok_low_k}
\end{figure}
%
%

Let us now identify which of the non-perturbative production mechanisms discussed above is dominant. 
For this purpose, we consider the two cases:

\begin{itemize}

\item{\underline{When $m_\chi < k \ll m_\phi$}: In this case the dimples in $\tilde{\omega}_k^2$, see Fig.~(\ref{fig:omega_k_no}), are almost completely on the negative side, in which case the tachyonic production is expected to be significant. Moreover, the parametric excitation in this case is expected to be insignificant as the violation of adiabaticity condition is expected to be moderate due to the smallness of the part of the dimple in $\tilde{\omega}_k^2$ residing on the positive side. In other words, for $m_\chi < k \ll m_\phi$ the non-perturbative production of $\chi$ particles mostly take place via tachyonic resonance. Fig.~(\ref{fig:rhok_low_k}) shows the evolution of $\rho_k$ in time for $m_\chi < k \ll m_\phi$, where the effect of expansion is ignored. }

\item{\underline{When $m_\chi < m_\phi < k$}: in this case parts of those dimples, on whose sides the adiabaticity condition is violated, reside on the positive side allowing a room for parametric particle production. 
One can easily see from Fig.~(\ref{fig:rhok_k_no}) that the non-perturbative production of $\chi$ particles in this case is way more efficient.}

\end{itemize}

Therefore, one may conclude that non-perturbative production of light ALPs is dominant in the momentum range $ m_\phi \lesssim k \lesssim m_\phi \Phi/ \sqrt{2} f_\chi$ with parametric resonance being the dominant production mechanism. This is also the case for Heavy ALPs, $m_\chi \gtrsim m_\phi$, which can be understood by comparing the $\rho_k$ for $k>m_\chi$, Fig.~(\ref{fig:rhok_k_no}), to that for $k<m_\chi$, Fig.~(\ref{fig:rhok_m_no}). Clearly, $\rho_k$ in the later case is way less than for the former.

\subsubsection{{\bf Non-perturbative excitations of ALPs, with expansion}}

\begin{figure}[ht]
\centering
\includegraphics[width=8.0cm]{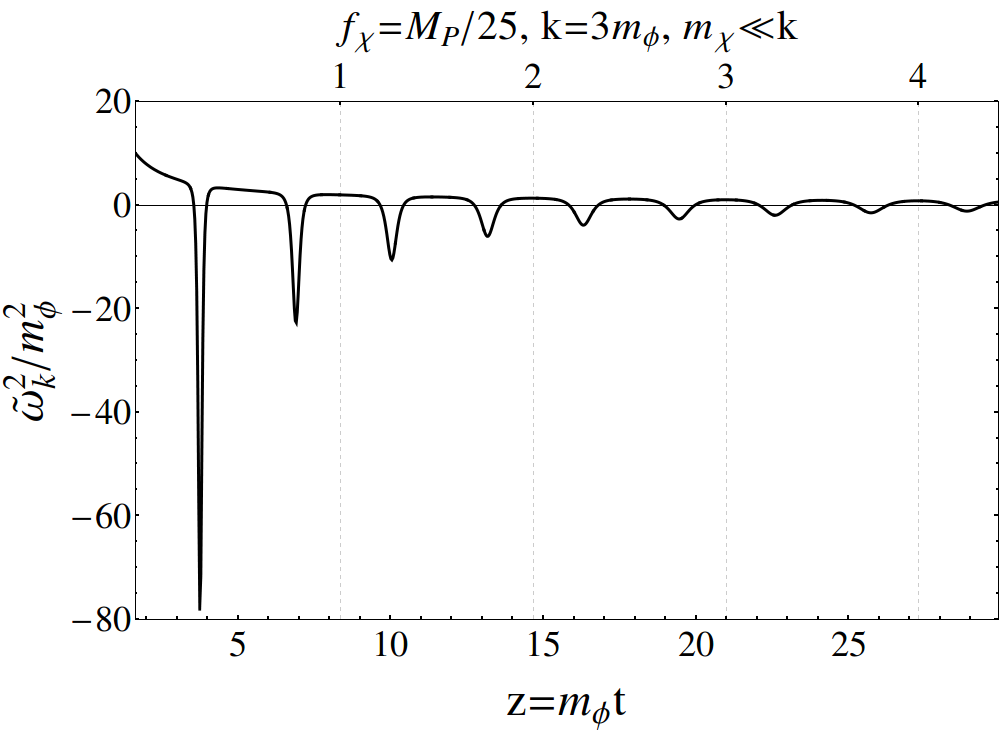}
\caption{Same as Fig.~(\ref{fig:omega_k_no}), but now the cosmic expansion is included. One can see that the tachyonic instabilities become dominant 
within few oscillations of $\phi$ field. The corresponding number of $\phi$ oscillations is shown on the upper horizontal axis. }
\label{fig:omega_k}
\end{figure}

\begin{figure}[ht]
\centering
\includegraphics[width=8.0cm]{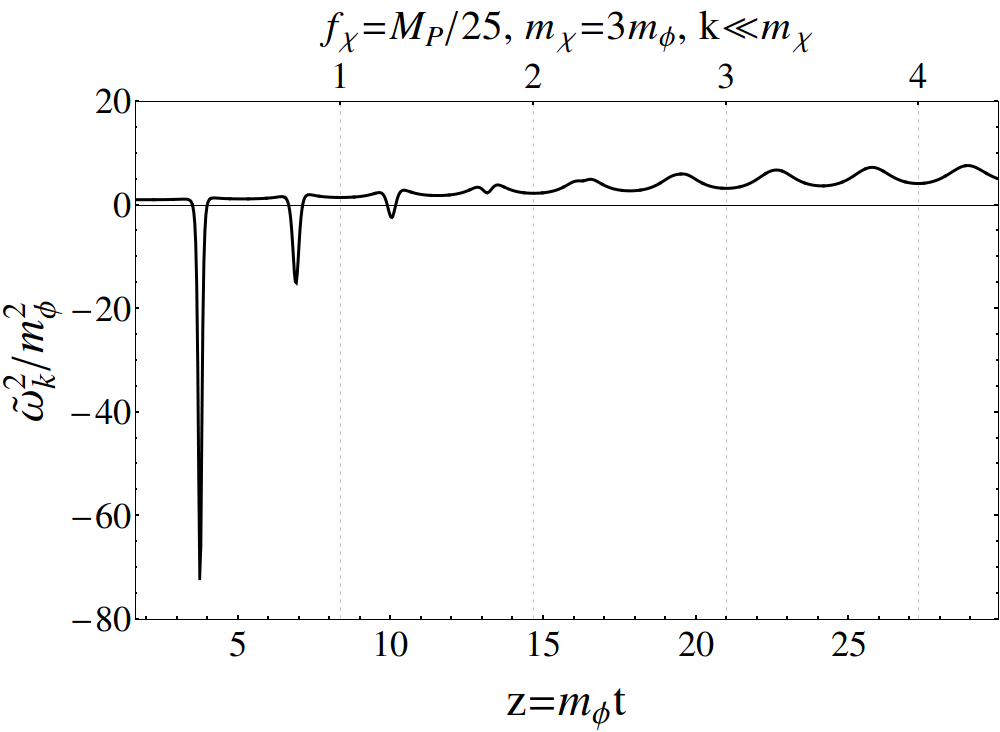}
\caption{Same as Fig.~(\ref{fig:omega_m_no}), but now cosmic expansion effect is included. One can see that the tachyonic instabilities become important within 
few oscillations of $\phi$. The corresponding number of $\phi$ oscillations is shown on the upper horizontal axis. }
\label{fig:omega_m}
\end{figure}

Let us now include the effect of expansion. This will result in lowering the production rate in two ways. First, the energy density of ALPs, $\rho_k$, with a given momenta dilutes with the expansion, which becomes obvious at late times where the excitation of $\chi_k$ quanta is less significant. Second, the amplitude of $\phi$ oscillations dies out with expansion, $\Phi \propto a^{-3/2}$, causing the tachyonic dimples in $\tilde{\omega}_k^2$ to become less and less deeper with time, see Figs.~(\ref{fig:omega_k}) and (\ref{fig:omega_m}). Therefore, one would expect that the non-perturbative excitation of $\chi$ quanta will seize to be significant after a few oscillations of $\phi$.\\

\paragraph{\centering {\bf Non-perturbative production of Light ALPs,  with $ m_\chi \ll m_\phi$}\\}
\noindent

For light ALPs with $ m_\chi \ll m_\phi$, we saw earlier that the non-perturbative production of ALPs is dominated by modes with momenta in the range
$m_\phi \lesssim  k \lesssim m_\phi \Phi/\sqrt{2} f_\chi$. Thus, except for the dimples in $\tilde{\omega}_k^2$, it behaves as $k^2/a^2$,
for which case it redshifts to lower values with the dimples keep being present, albeit becoming shallower, see Fig.~(\ref{fig:omega_k}); as a result the violation of adiabaticity due to the sharp change in $\tilde{\omega}_k^2$ on the sides of those dimples becomes more and more moderate with time.
Hence depending on the scale $f_\chi$, the non-perturbative excitations of $\chi$ quanta will become less and less efficient with time, until it is completely taken over by the expansion of the Universe. Nevertheless, as we shall see, we would end up exciting a large abundance of ALPs even in this case.

%
%
%
%
%
%
%
\begin{figure}[ht]
\centering
\includegraphics[width=8.0cm]{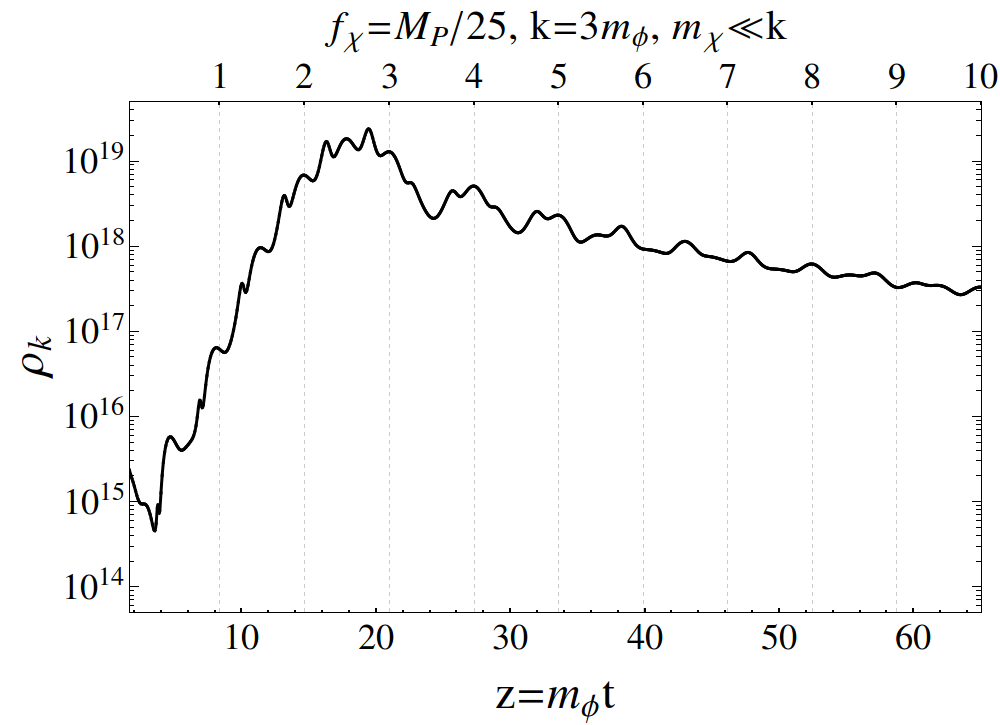}
\caption{The above figure shows the evolution of  $\rho_k$ in time, where the cosmic  expansion is included. Here  time is measured in units $m_\phi^{-1}$; the corresponding number of $\phi$ oscillations is shown on the upper horizontal axis.}
\label{fig:rhok_k}
\end{figure}
%
%

In Fig.~(\ref{fig:rhok_k}), we have shown the evolution of $\rho_k$ for $f_\chi=\Phi_{\rm I}/25$, $m_\chi \ll m_\phi$ and $k=3m_\phi$, by taking into account the effect of expansion. One can see that the non-perturbative production of ALPs becomes insignificant roughly at the end of the third oscillation of $\phi$.

\begin{figure}[ht]
\centering
\includegraphics[width=8.0cm]{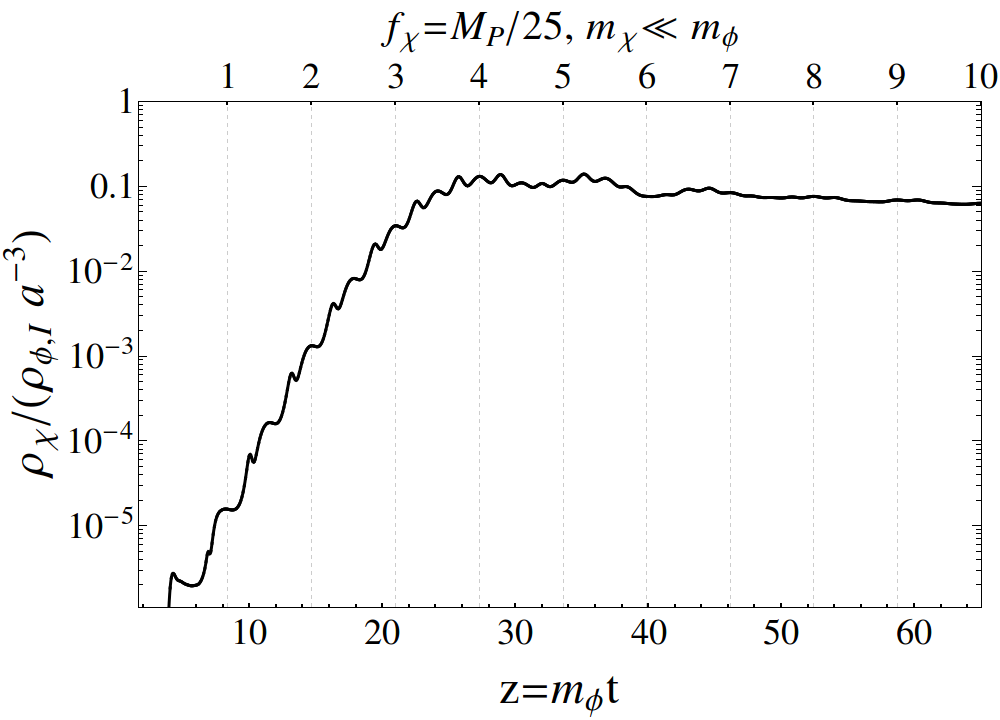}
\caption{The above figure shows the evolution of  $\rho_\chi/\rho_{\phi,I}$ in time  for $m_\chi \ll m_\phi$, where $m_\phi=10^{13}$ GeV. Here time is measured in units $m_\phi^{-1}$; the corresponding number of $\phi$ oscillations is shown on the upper horizontal axis.  This analysis includes cosmic expansion.}
\label{fig:rho_light}
\end{figure}

Next, we calculate
the energy density of the $\chi$ particles, defined by Eq.~(\ref{eq:rhochi}).
For $f_\chi=\Phi_{\rm I}/25$ and $m_\chi \ll k$, we have shown the time evolution of
$\rho_\chi/(\rho_{\phi,I} a^{-3})$ in Fig.~(\ref{fig:rho_light}).
In this case roughly $10\%$ of the condensate's energy get gets transferred to the excitations of the $\chi$ field.  However, when $f_\chi < \Phi_{\rm I}/27$, most of the energy stored in the $\phi$ field can be drained in which case the problem becomes nonlinear.\\
%
%
%
%
%

\paragraph{\centering {\bf Non-perturbative production of heavy ALPs, with $ m_\chi \gtrsim m_\phi$}\\}
\noindent

let us now consider the effect of expansion on the non-perturbative production of heavy ALPs. We saw that the non-pertrbative excitation of modes with $k,m_\phi<m_\chi$ is way less efficient than for modes with $m_\phi< m_\chi < k$, see Figs.~(\ref{fig:rhok_k_no}) and (\ref{fig:rhok_m_no}). Including the expansion effect, the difference in efficiency between the two cases becomes even more pronounced for the following reason. As the amplitude of $\phi$ oscillation decays with time, $\tilde{\omega}_k^2$ will have different behaviour for $k > m_\chi$ and $k <m_\chi$. For the former case, $\tilde{\omega}_k^2$ tends toward smaller values with the dimples in $\tilde{\omega}_k^2$ keep being present, albeit becoming shallower and the violation of adiabaticity condition on the sides of those dimples becomes milder with time, see Fig~(\ref{fig:omega_k}).

%
%
%
%
%
\begin{figure}[ht]
\centering
\includegraphics[width=8.0cm]{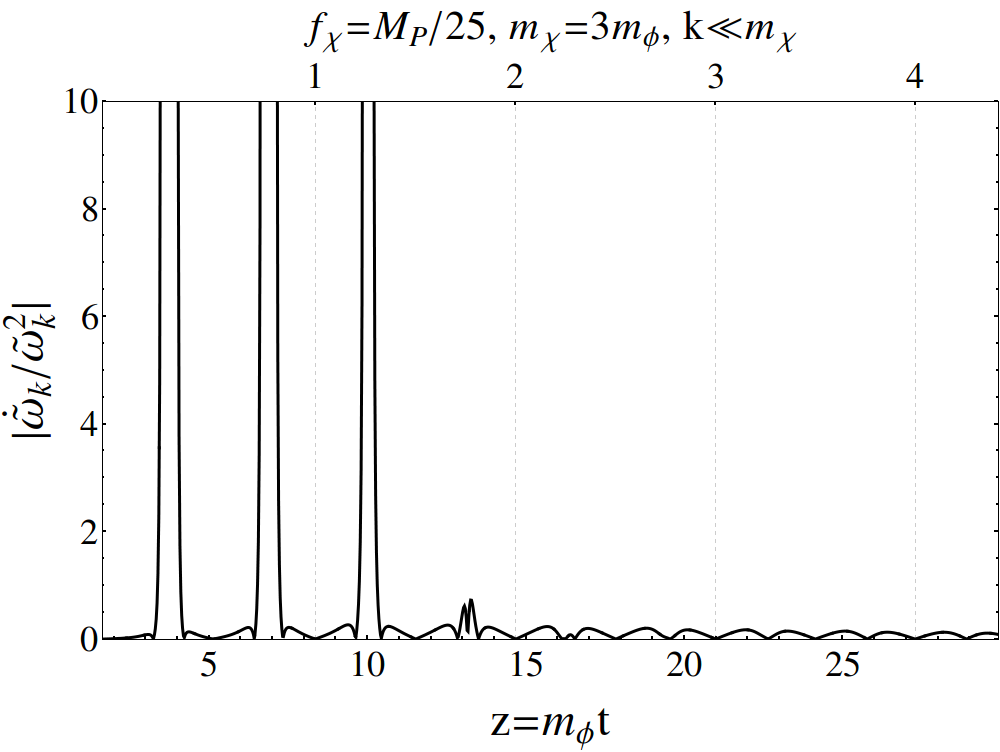}
\caption{Same as Fig.~(\ref{fig:adiab_m_no}), but now the cosmic expansion is taken into account. The evolution of $\tilde{\omega}_k$ seizes to be non-adiabatic after a few oscillations of $\phi$. The corresponding number of $\phi$ oscillations is shown on the upper horizontal axis. }
\label{fig:adiab_m}
\end{figure}
%
%
%
\begin{figure}[ht]
\centering
\includegraphics[width=8.0cm]{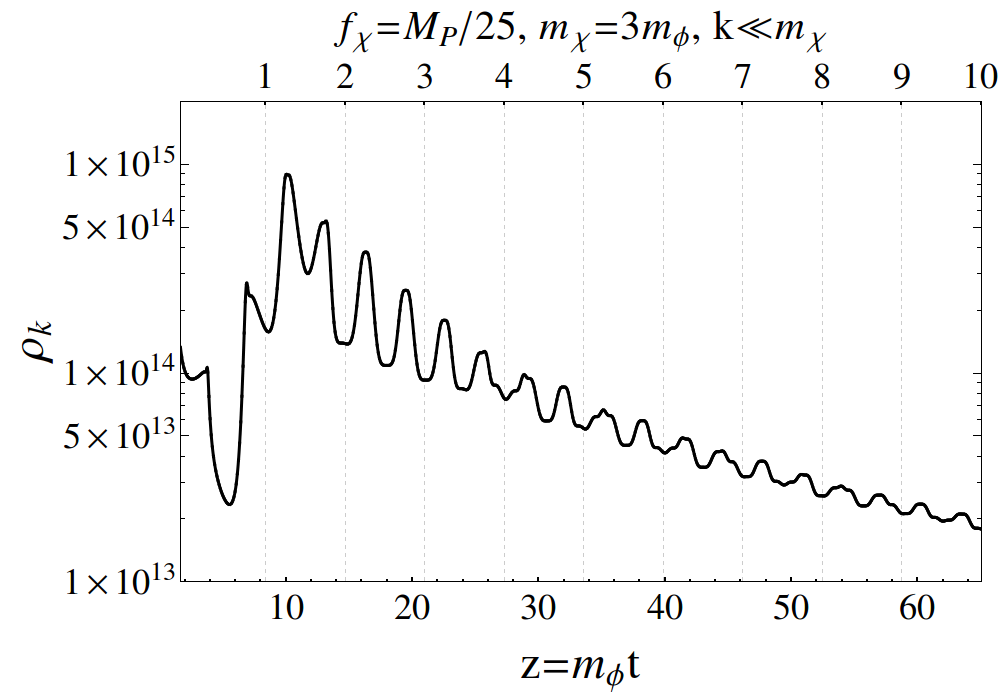}
\caption{The above figure shows the time evolution of  $\rho_k$. As expected, the excitation of such heavy ALPs takes place during the first  few oscillations of $\phi$. Here time is measured in units $m_\phi^{-1}$; the corresponding number of $\phi$ oscillations is shown on the upper horizontal axis.  This analysis includes cosmic expansion.}
\label{fig:rhok_m}
\end{figure}
%
%

On the other hand, for $k <m_\chi$, $\tilde{\omega}_k^2$ tends toward larger positive values, i.e toward its maximum value which is $m_\chi^2$, with the dimples becoming shallower with time and even disappearing in a couple oscillations of $\phi$, see Fig~(\ref{fig:omega_m}). This behaviour can be also seen from Fig.~(\ref{fig:omega_alpha_2}). With the disappearance of the dimples in $\tilde{\omega}_k^2$, the adiabaticity condition is no more violated which can be seen from Fig.~(\ref{fig:adiab_m}).
Therefore, one would expect that the non-perturbative excitation of modes with $k>m_\chi$ last longer that that for modes with $k<m_\chi$.

This render the non-perturbative particle production in the later case less efficient which can be also seen from Figs.~(\ref{fig:rhok_k}) and (\ref{fig:rhok_m}) that show the time evolution of $\rho_k$ for $ k>m_\chi$ and $ k<m_\chi$, respectively, where $f_\chi= \Phi_{\rm I}/25$. For the later case, the particle production seizes to be efficient roughly after the first $\phi$ oscillation, whereas it keeps being 
significant roughly till the end of the third $\phi$ oscillation for the former case.
%
%
\begin{figure}[ht]
\centering
\includegraphics[width=8.0cm]{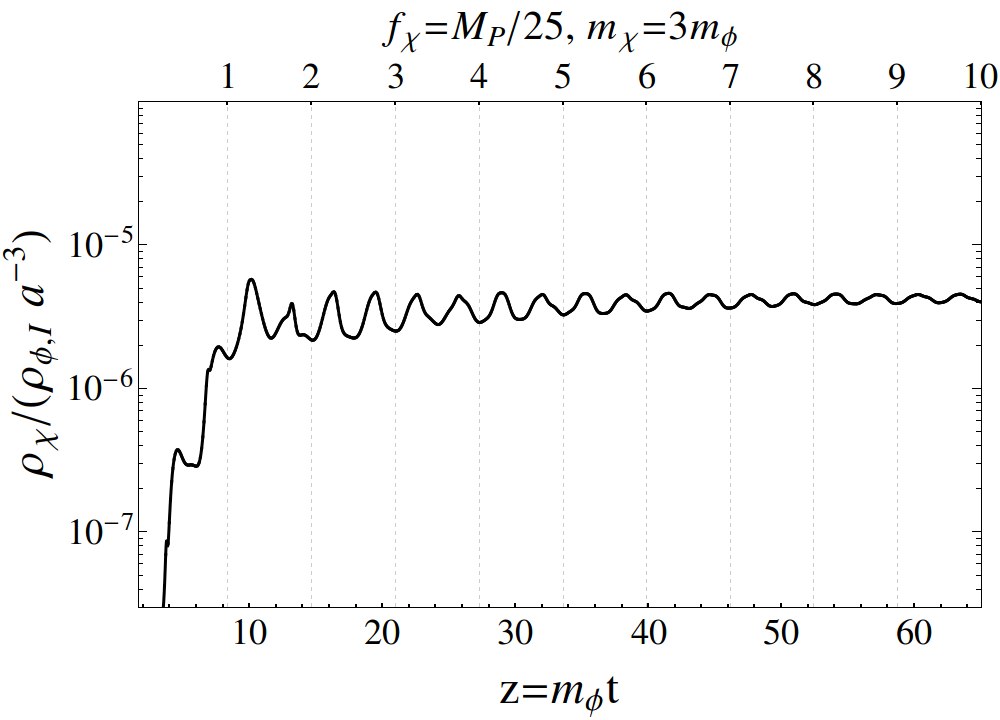}
\caption{The above figure shows the time evolution of  $\rho_\chi/\rho_{\phi,I}$ for $m_\chi = 3 m_\phi$, where $m_\phi=10^{13}$ GeV. Here time is measured in units $m_\phi^{-1}$; the corresponding number of $\phi$ oscillations is shown on the upper horizontal axis.  This analysis includes cosmic expansion.}
\label{fig:rho_heavy}
\end{figure}
%
%
%
%
\begin{figure}[ht]
\centering
\includegraphics[width=8.0cm]{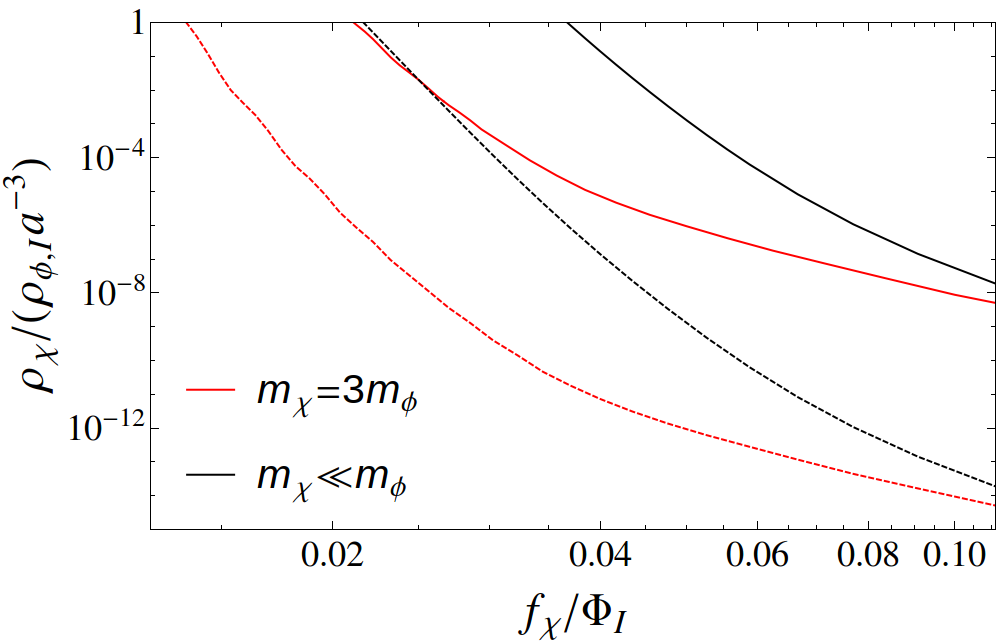}
\caption{The above figure shows $\rho_\chi/\rho_{\phi,I}$ as a function of ratio $f_\chi/\Phi_{\rm I}$ after $5$ $\phi$ oscillations for $m_\phi=10^{10}$ GeV (dashed curves) and $m_\phi=10^{13}$ GeV (solid curves). The black (red) curve corresponds to light (heavy) ALPs, respectively. This analysis includes cosmic expansion.}
\label{fig:rho_fchi}
\end{figure}
%
%

Now we proceed to calculate the $\rho_\chi$ for $m_\chi > m_\phi$ taking into account the effect of expansion. In Fig.~(\ref{fig:rho_heavy}), we have shown the relative energy density of ALP to that of $\phi$ particles, respectively.
It is evident that particle creation is way less efficient for heavy ALPs as compared to light ones. This is because of the smallness of the window for the more efficient non-perturbative production of modes with momenta $k>m_\chi$ (i.e $m_\chi < k \lesssim m_\phi \Phi/\sqrt{2} f_\chi $ with $m_\chi \gtrsim m_\phi$) in contrast to the case
$m_\chi < m_\phi$.

Therefore, in order for the non-perturbative production of heavy ALPs to be efficient, a smaller $f_\chi/\Phi_{\rm I}$ is required. 
For instance, when $f_\chi \geq \Phi_{\rm I}/47$  and $m_\chi =3m_\phi$, most the condensate's energy density can be transferred to the $\chi$ field, within 
few oscillations of the scalar condensate, $\phi$. Heavier ALPs requires 
even smaller $f_\chi \ll \Phi_{\rm I}$. In Fig.~(\ref{fig:rho_fchi}), we have shown a plot of the energy density of $\chi$ field to the redshifted initial energy density of the scalar condensate as a function of the ratio $f_\chi/\Phi_{\rm I}$ after $5$ $\phi$ oscillations. The black (red) curve corresponds to light (heavy) ALPs, respectively. 
One can easily see the difference in efficiency between the two cases. It is also clear that the ratio $f_\chi/\Phi_{\rm I}$ being $\ll 1$ is crucial for efficient particle creation including
cosmic expansion.

\section{Conclusion} \label{sec:conclusion}

In this paper, we have considered the non-perturbative production of ALPs during the coherent oscillations of 
a scalar condensate with mass $m_\phi$. The ALPs couple only derivatively to other fields including
inflaton or any moduli field. We focused on the couplings, dimensional $5$: $\phi (\partial \chi)^2/f_\chi$, and 
dimensional $6$ operators: $\phi^2 (\partial \chi)^2/f_\chi^2 $.

In both cases, when $f_\chi \gtrsim \Phi_{\rm I} $, there is no significant non-perturbative production of ALPs.
Nevertheless, when $f_\chi \ll \Phi_{\rm I} $, the non-perturbative excitation of ALPs is significant.
However, in case of dimensional $5$ operator, $\phi (\partial \chi)^2/f_\chi$, singularities in both the damping coefficient, 
$\alpha$ and the frequency squared, $\omega_k^2$ arise when $\phi = -f_\chi$. These singular 
points occur when the kinetic term changes from being positive to negative and vice versa giving 
rise to an unphysical ghost-like behaviour of $\chi$ between these singular points, i.e. when $\phi <-f_\chi$.

The energy of the scalar condensate can still be drained even before the dynamical system hits the first singular point rendering the problem 
non-linear before the first singular point. This will be the case if any coupling with odd powers of $\phi$ dominates the interaction with ALPs.

On the contrary, such ghost-like behaviour does not arise in even dimension interactions, $\phi^2 (\partial \chi)^2/f_\chi^2 $.
In this case, ALPs with masses in the range $0 - m_\phi \Phi/\sqrt{2} f_\chi$ can be copiously produced via non-perturbative effects, provided that $ f_\chi \ll \Phi_{\rm I}$.
If $f_\chi$ is sufficiently below, $\Phi_{\rm I}$ ($f_\chi < \Phi_{\rm I}/27$ for $m_\chi \ll m_\phi$ and even smaller values for heavy ALPs, $m_\chi \gtrsim m_\phi$, for instance $f_\chi \lesssim \Phi_{\rm I}/ 47$ for $m_\chi = 3 m_\phi$), most of the energy density of the scalar condensate would be transferred to the $\chi$ excitations. Irrespective of inflaton or moduli as a scalar condensate,
this will definitely cause an alarming problem for ALP-dominated Universe, as it would lead to significant departure from the Standard Model cosmology and could spoil the success of BBN, unless the ALP is unstable and decay before the era of BBN. These constraints will be discussed in a separate publication.

Moreover, if $\chi$ is stable enough to be the dark matter, one can show that ALPs with a wide mass range can
overclose the universe. Only in a very fine-tuned region of the parameter space can one match the observed DM abundance.
Furthermore, the recent Planck constraint on the number of relativistic species, $N_{\rm eff} =3.15^{+0.41}_{-0.40} $~\cite{Planck:2015xua}, 
which places an upper bound on the abundance of dark radiation, $\Omega_{\rm dark} h^2\leq 2.92\times 10^{-6}$, can be used to further constrain 
the parameter space toward the lower mass end~\cite{Cicoli:2012aq,Dev:2013yza}.

In principle, ALPs could also couple to other fields. For instance, it could couple to the Higgs via the operator:
$\partial_\mu \partial^\mu \chi \Phi^\dagger \Phi/f_\chi$, or to  a gauge boson via the operator:
$\chi F_{\mu \nu} \tilde{F}^{\mu \nu}/f_\chi$,
which in the ALP's rest frame leads to a decay rate: $\Gamma \sim  m_\chi^3/f_\chi^2$.
The ALPs could also couple to fermionic fields via the operator:
$\partial^\mu \chi \bar{\psi} \gamma_\mu \gamma_5 \psi /f_\chi$.
The decay rate of ALPs will be then given by: $ \Gamma \sim m_\psi^2 m_\chi/f_\chi^2$, in the $\chi$ rest frame.
The perturbative decay rates above,
being already slow,
are even slower for lighter ALPs, $m_\chi \ll m_\phi$, which are relativistic, due to the smallness of their mass and the time dilation effect when the decay rates are transformed from the particles rest frame to the cosmic comoving frame.

This can be used to further constrain the ALP's parameter space. For instance, if ALP constitute a component or the whole of DM, their mean life time should be longer than the age of the universe which places an upper bound of $40~{\rm MeV}~({f_\chi}/{10^{18} {\rm GeV}}) ^{2/3} $ on the mass for ALPs. Heavier ALPs would decay at earlier stages and hence, their mean life time and consequently, their mass would be subject to several constraints such as from CMB, BBN, and other astrophysical observations.

In view of our results, it would be indeed interesting to revisit some of the earlier computations on axion production in the string theory context, where most of the analysis was done in a perturbative decay of the inflaton/moduli into axions~\cite{Cicoli:2010ha,Cicoli:2010yj,Cicoli:2012aq,Higaki:2012ar}. Our current study highlights that by taking into account of various non-perturbative effects, the Universe after string theory motivated inflationary models could be filled abundantly with non-thermal, both heavy and light axions. Although, we have discussed the ALP production for a single scalar condensate, with multi-scalar condensate, the situation may get even worst, such as in the case of {\it assisted inflation}~\cite{Liddle:1998jc}.

In future, it would be also interesting to introduce, systematically,  higher order interactions and sum them up in a natural way such 
that the theory of axion interaction becomes  an {\it infinite derivative} theory~\cite{Chialva:2014rla}, very similar to the gravitational interaction~\cite{Biswas:2005qr}.

Before we conclude, the most important aspect of our analysis suggests that the ALPs can be overproduced in the early Universe for $f_\chi \ll \Phi_{\rm I}$ during
the coherent oscillations of the inflaton or a moduli condensate, the summary plot is Fig.~\ref{fig:rho_fchi}, which shows how the energy density in the ALPs for 
a wide range of masses can overclose the Universe for $f_\chi/\Phi_{\rm I} <1$.  In past much of the attention has been paid to understand perturbative production of ALPs,
but the non-perturbative production of ALPs is perhaps the most dangerous channel to drain the inflaton or moduli energy density. In light of this discussion, it is important
that we reconsider some of the analysis of axion production in the early Universe.

\section*{acknowledgemnets}

The authors would like to thank Bhupal Dev for initial collaboration, we would also like to thank 
David Cerdeno, Steve Abel,  Masahide Yamaguchi, Robert Brandenberger and Celine Boehm for discussions.
AM is supported by the STFC grant ST/J000418/1, and the JSPS visiting scientist's award. SQ is funded by the King Abdulaziz University.
AM acknowledges the kind hospitality from Tokyo Institute for Technology, Japan, and IPPP, Durham, 
during the course of this work.\\

\appendix

\section{ALP's energy density} \label{app:nk}

The stress-energy tensor for  the Lagrangian under consideration, Eq.~({\ref{lag1}}), is given by
\begin{eqnarray}
T_{\mu \nu} &=& 
\left( 1+ \frac{ \phi}{f_\chi} + \frac{\phi^2}{2 f_\chi^2}\right) \partial_\mu \chi \partial_\nu \chi + \partial_\mu \phi \partial_\nu \phi
\nonumber \\
&-&  \frac{1}{2} g_{\mu \nu} g^{\alpha \beta} \left( 1+ \frac{\phi}{f_\chi} + \frac{ \phi^2}{2 f_\chi^2}\right) \partial_\alpha \chi \partial_\beta \chi
\nonumber \\
&-&  \frac{1}{2} g_{\mu \nu} g^{\alpha \beta} \partial_\alpha \phi \partial_\beta \phi
+ \frac{1}{2} g_{\mu \nu} \left(  m_\chi^2 \chi^2 + m_\phi^2 \phi^2 \right)
\, .
\nonumber \\
\end{eqnarray}
Hence, the energy density of the $\chi$ field is then given by
\begin{eqnarray}
\rho_\chi &=&
\frac{1}{2} \left( 1+ \frac{\phi}{f_\chi} + \frac{\phi^2}{2 f_\chi^2}\right)
\left[ \vert \dot{\chi} \vert^2 + \frac{1}{a^2} \vert \nabla \chi \vert^2\right]
 + \frac{1}{2} m_\chi^2 \vert \chi \vert^2 \, ,
 \nonumber \\
\end{eqnarray}
where the prefactor $\left( 1+ \phi/f_\chi + \phi^2/ 2 f_\chi^2 \right)$ exhibits an oscillating behaviour with amplitude decaying  with the expansion of the Universe. When  the amplitude of $\phi$ oscillation, $\Phi$, drops below $f_\chi$, this prefactor can be ignored. Furthermore, the energy density for a given momentum mode would then be given by
\begin{eqnarray} \label{eq:rhok}
\rho_{k} &=&
 \left( 1+ \frac{\phi}{f_\chi} + \frac{\phi^2}{2 f_\chi^2}\right) 
\left[ \frac{1}{2} \vert \dot{\chi}_k \vert^2 + \frac{1}{2} \omega_{k,eff}^2 \vert \chi_k \vert^2 \right]
\, ,
\nonumber \\
\end{eqnarray}
where $\omega_{k,eff}$, denoting the effective frequency of $\chi_k$ oscillation, is defined by
\begin{eqnarray}
\omega_{k,eff} &\equiv 
\left( \frac{k^2}{a^2} 
  +  \frac{m_\chi^2 }{ 1+ \frac{\phi}{f_\chi} + \frac{\phi^2}{2 f_\chi^2} }
  \right)^{1/2}
\, .
\end{eqnarray}
For simplicity of the notation, we will call $\omega_{k,eff}$, just $\omega_{k}$. In terms of $\rho_k$, the ALP's energy density is given by
\begin{eqnarray} \label{eq:rhochi}
\rho_\chi =
\frac{1}{2 \pi^2}  \int dk ~ k^2   \rho_k
\, .
\end{eqnarray}
%
%

\section{Canonically normalised Lagrangian} \label{app:canonocal}

Instead of our approach, one could have first  canonically 
normalized the kinetic term in the Lagrangian Eq.~(\ref{lag1}), 
which could be easily achieved by 
following field redefinition: $$\chi \rightarrow (1+\phi/f_\chi + \phi^2/f_\chi^2)^{-1/2} \chi. $$ 
The Lagrangian then reads
\begin{widetext}
\begin{equation}
{\cal L}  =
 \frac{1}{2}\partial^\mu \phi \partial_\mu \phi  - \frac{1}{2} m_\phi^2 \phi^2
+ \frac{1}{2}\partial^\mu \chi \partial_\mu \chi  - \frac{1}{2} m_\chi^2 \chi^2
+  \frac{1}{2} \frac{m_\chi^2 (\frac{\phi}{f_\chi} + \frac{\phi^2}{2f_\chi^2}) \chi^2}{(1+\frac{\phi}{f_\chi} + \frac{\phi^2}{2f_\chi^2})}
- \frac{1}{2} \frac{(\frac{\partial_\mu \phi}{f_\chi} + \frac{\phi \partial_\mu \phi}{f_\chi^2})}{(1+\frac{\phi}{f_\chi} + \frac{\phi^2}{2f_\chi^2})} \chi \partial^\mu \chi
+ \frac{1}{8} \frac{(\frac{\partial_\mu \phi}{f_\chi} + \frac{\phi \partial_\mu \phi}{f_\chi^2})}{(1+\frac{\phi}{f_\chi} + \frac{\phi^2}{2f_\chi^2})^2} (\frac{\partial^\mu \phi}{f_\chi} + \frac{\phi \partial^\mu \phi}{f_\chi^2}) \chi^2
 \, .
\label{lag2}
\end{equation} 
\end{widetext}
Assuming that $\phi$ remains a homogeneous condensate, one can then derive the following equation of motion for the $\chi$ field:
\begin{widetext}
\begin{equation}
\label{eq:chieomp}
\partial^\mu \partial_\mu \chi + 3 H \partial_0 \chi
+\left[ 
\frac{1}{4}  \frac{(\frac{\partial_0 \phi}{f_\chi} + \frac{\phi \partial_0\phi}{f_\chi^2})
(\frac{\partial^0 \phi}{f_\chi} + \frac{\phi \partial^0\phi}{f_\chi^2})}{(1+\frac{\phi}{f_\chi}+ \frac{\phi^2}{2 f_\chi^2})^2}
+ \frac{m_\chi^2 - \frac{1}{2}(\frac{\partial_0\partial^0 \phi}{f_\chi} + \frac{\phi \partial_0\partial^0 \phi}{f_\chi^2} + \frac{\partial_0 \phi \partial^0 \phi }{f_\chi^2}) - \frac{3 H}{2} (\frac{\partial_0 \phi}{f_\chi} + \frac{\phi \partial_0 \phi}{f_\chi^2})} {(1+\frac{\phi}{f_\chi} + \frac{\phi^2}{2 f_\chi^2})}
\right] \chi = 0
\, .
\end{equation}
\end{widetext}
Expanding $\chi$ in terms of the momentum modes, Eq.~(\ref{eq:chieomp}) can be rewritten as
\begin{eqnarray} \label{eq:chikeomp}
\ddot{\chi}_k +  3 H(t) \dot{\chi}_k + \omega_k^2(t) \chi_k   =0 \, ,
\end{eqnarray}
where
\begin{eqnarray} \label{eq:omegap}
 \omega_k^2(t) &=&
\frac{k^2}{a^2}  +
\frac{ m_\chi^2 - \frac{1}{2} (\frac{\ddot{\phi}}{f_\chi} + \frac{\phi \ddot{\phi}}{f_\chi^2}
+ \frac{\dot{\phi}^2}{f_\chi^2}) 
- \frac{3 H}{2} (\frac{\dot{\phi}}{f_\chi} 
+ \frac{\phi \dot{\phi}}{f_\chi^2}) }
 {(1+\frac{\phi}{f_\chi} 
 + \frac{\phi^2}{2 f_\chi^2})} 
\nonumber \\
&&
+\frac{1}{4}  \frac{(\frac{\dot{\phi}}{f_\chi} + \frac{\phi \dot{\phi}}{f_\chi^2})
(\frac{\dot{\phi}}{f_\chi} + \frac{\phi \dot{\phi}}{f_\chi^2})}{(1+\frac{\phi}{f_\chi}+ \frac{\phi^2}{2 f_\chi^2})^2} \,.
\end{eqnarray}
When the interaction with the scalar condensate is dominated by the dimensional 5 coupling, Eq.~(\ref{eq:omegap}) can be approximated by
\begin{eqnarray} \label{eq:omegap5}
 \omega_k^2(t) \simeq
\frac{k^2}{a^2}  +
\frac{ m_\chi^2 - \frac{\ddot{\phi}}{2f_\chi}  
-  \frac{3 H}{2f_\chi} \dot{\phi} }
 {(1+\frac{\phi}{f_\chi})} 
+\frac{(\frac{\dot{\phi}}{2 f_\chi})^2}{(1+\frac{\phi}{f_\chi})^2} \,.
\end{eqnarray}
On the other hand, when the dimension 6 operator dominates the interaction with the scalar condensate, Eq.~(\ref{eq:omegap}) can be approximated by
\begin{eqnarray} \label{eq:omegap6}
 \omega_k^2(t) &\simeq &
\frac{k^2}{a^2}  +
\frac{ m_\chi^2 - \frac{1}{2} (\frac{\phi \ddot{\phi}}{f_\chi^2}
+ \frac{\dot{\phi}^2}{f_\chi^2}) 
-  \frac{3 H }{2f_\chi^2} \phi \dot{\phi}}
 {(1+ \frac{\phi^2}{2 f_\chi^2})} 
+ \frac{( \frac{\phi \dot{\phi}}{2f_\chi^2})^2}{(1 + \frac{\phi^2}{2 f_\chi^2})^2}
\,.
\nonumber \\
\end{eqnarray}
Eq.~(\ref{eq:chikeomp}) with frequency $\omega_k$ defined by Eq.~({\ref{eq:omegap6}}) for the dimensional 6 operator becomes exactly the same as Eq.~(\ref{eq:ykeom2}) with $\tilde{\omega}_k$ defined by Eq.~(\ref{eq:ykomega22}), along with  $z=m_\phi t$. One can also show the same happens for dimensional 5  and all higher order operators.
Both the approaches discussed in this paper, i.e. with and without canonical kinetic term, would yield similar numerical result for the energy density, $\rho_\chi$, stored 
in the ALPs. This has been numerically verified.


\end{document}